\documentclass[journal]{IEEEtran}

\usepackage{amsmath,amssymb,amsfonts, amsthm}
\usepackage{array}
\usepackage[caption=false,font=footnotesize,labelfont=sf,textfont=sf]{subfig}
\usepackage{textcomp}
\usepackage{stfloats}
\usepackage{url}
\usepackage{verbatim}
\usepackage{graphicx}
\usepackage{cite}

\usepackage[dvipsnames]{xcolor}
\usepackage[linesnumbered,ruled,vlined]{algorithm2e}
\SetNoFillComment
\usepackage[noend]{algpseudocode}
\usepackage{multirow}
\usepackage{tikz}
\usepackage{pifont}
\usepackage{arydshln}
\usepackage{orcidlink}
\usepackage{booktabs}
\usepackage{bigdelim} 
\usepackage{makecell}
\usepackage{breqn}
\usepackage{adjustbox}
\usepackage{enumerate}
\usepackage{bm}

\newcommand{\ciphr}{{\sffamily\textbf{CIPHR}}}
\newcommand{\ciphrs}{\textbf{\ciphr~}}

\newcommand{\cmark}{\ding{51}}%
\newcommand{\xmark}{\ding{55}}%
\newcommand{\redx}{\textcolor{OrangeRed}{\xmark}}
\newcommand{\grnc}{\textcolor{ForestGreen}{\cmark}}
\newcommand{\textbu}[1]{{\sffamily\textbf{\underline{#1}}}}
\newcommand{\ddg}{\textbf{\textsuperscript{$\ddagger$}}}
\newcommand{\cdt}{\textbf{\textsuperscript{$\circledast$}}}
\newcommand{\bstars}{\textbf{\textsuperscript{$\bigstar$}}}
\newcommand{\cluts}{$\{\mathbf{CL2,CL3,CL4}\}$}
\newcommand{\bits}{$\mathbf{\left|Bitstream\right|}$}
\newcommand{\ho}{\hspace{1pt}}

\begin{document}
\title{CIPHR: Cryptography Inspired IP Protection through Fine-Grain Hardware Redaction}

\author{Aritra~Dasgupta \orcidlink{0000-0002-0786-9185},~\IEEEmembership{Graduate Student Member,~IEEE}, Sudipta~Paria \orcidlink{0009-0002-7726-8032},~\IEEEmembership{Graduate Student Member,~IEEE}, and~Swarup~Bhunia \orcidlink{0000-0001-6082-6961},~\IEEEmembership{Fellow,~IEEE}
\thanks{A. Dasgupta, S. Paria, and S. Bhunia are with the Department of Electrical and Computer Engineering, University of Florida, Gainesville, FL, 32611 USA (e-mail: aritradasgupta@ufl.edu, sudiptaparia@ufl.edu, swarup@ece.ufl.edu).}
}



\maketitle

\begin{abstract}
Hardware intellectual property (IP) in the globalized integrated circuit (IC) supply chain is exposed to a wide range of confidentiality and integrity attacks by untrusted third-party entities. Existing IP-level countermeasures, such as logic locking, hardware obfuscation, camouflaging, and redaction, have aimed at addressing these them. In particular, hardware redaction has emerged as a robust approach for IP protection against confidentiality attacks, including reverse engineering. We note that existing IP protection approaches, including the ones based on hardware redaction, tend to leave behind structural artifacts that can be exploited by adversaries to bypass protections or predict unlocking keys, using the knowledge of known designs, akin to a known-plaintext attack (KPA) in cryptography. In this work, we present \ciphr, a robust fine-grain hardware redaction methodology inspired by the cryptographic property of indistinguishability. The proposed approach utilizes novel heuristic-driven randomization to introduce significant structural transformations into the redacted designs. We employ structural analysis metrics to evaluate the security achieved by \ciphrs compared to various state-of-the-art IP protection techniques. Multiple open-source benchmark designs are used to demonstrate that fine-grain redaction in \ciphrs is robust, scalable, and indistinguishable against structural attacks. 
\end{abstract}


\section{Introduction}
\label{sec:intro}

\begin{figure}[!htbp]
    \centering
    \includegraphics[width=\columnwidth]{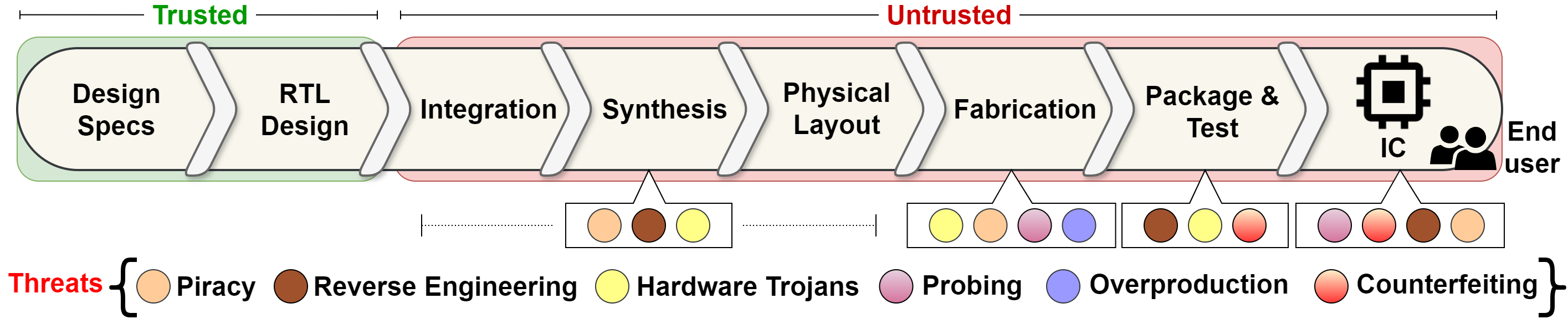}
     \caption{Major threats to IP confidentiality \& integrity throughout its life cycle. Under the zero-trust model for hardware security, no entity in the supply chain can be fully trusted, apart from the IP vendor/owner/designer.}
    \label{fig:ic_design_flow}
\end{figure}

The growing demand for integrated circuits (ICs) and the globalization of the semiconductor manufacturing process have led to outsourcing critical stages of semiconductor design to untrusted overseas facilities. This necessitates a zero-trust security model, where no entity is considered fully trusted during the hardware intellectual property (IP) lifecycle. This makes the IPs vulnerable to various threats such as piracy, counterfeiting, reverse engineering (RE), and hardware Trojans, as shown in Fig. \ref{fig:ic_design_flow}, that can compromise sensitive information \cite{hw_security_book_2018_Bhunia}, emphasizing the need for robust protection of IP confidentiality and integrity. Logic Locking \cite{EPIC_2008, HARPOON_TCAD2009_Chakraborty} emerged as one of the prominent IP protection methodologies where additional key inputs are inserted to protect against RE. However, it is vulnerable to key extraction techniques that rely on Boolean satisfiability (SAT) solvers \cite{SAT_2015, AppSAT_HOST2017_Shamsi}. Other techniques \cite{AntiSAT_TCAD2019_Yang, SFLL_2017,SARO_2021,united_we_protect} that are provably secure against SAT-based attacks have been developed, but they remain susceptible to attacks \cite{SAIL_2018, SWEEP_2019,gnnunlock} involving structural analysis or machine learning. IC Camouflaging \cite{Camo_ICCAD2019_Li, split_manu}, designed to prevent RE, faces challenges such as high overhead, resource demands, limited flexibility, and vulnerability to advanced attacks.

Unlike traditional locking and camouflaging techniques, hardware IP redaction involves replacing security-critical logic blocks using programmable components, ensuring robust protection against unauthorized access or RE attacks. Coarse-grain redaction techniques utilize embedded FPGAs (eFPGA) \cite{eFPGA-CMU-2_2021, eFPGA-NYU_2021, ALICE_2022, SheLL_2023} that allow designers to implement security-critical modules within the eFPGA fabric \cite{OpenFPGA_2019, FABulous_2021}. Essentially, certain parts of the design are replaced with programmable eFPGA fabric, where the bitstream serves as a secret key. The adversary must fully reconstruct the bitstream in order for the entire system to function correctly. The eFPGA-based redaction techniques have demonstrated resistance to SAT-based attacks \cite{SAT_2015, AppSAT_HOST2017_Shamsi} but introduce substantial area overhead, making them impractical for many real-world applications. LUT-based obfuscation \cite{LUT-Lock_2018, Custom-LUT_2019} provides another alternative that utilizes multiplexers and lookup tables (LUTs) to replace critical logic to protect against RE attacks. \textit{EvoLUTe} \cite{EvoLUTe_2023} presents a fine-grain redaction technique that efficiently identifies and replaces sensitive logic cones with LUTs, offering some resilience to SAT-based attacks \cite{SAT_2015, AppSAT_HOST2017_Shamsi}. 

Although both coarse-grain and fine-grain redaction techniques show promise, they still suffer from significant power, performance, and area (PPA) overheads and offer limited flexibility. Furthermore, these techniques are vulnerable to more advanced structural attacks such as \cite{DANA_2020,gnnunlock,dasgupta2025latte,knowledge_guided_attack} due to limited structural randomization after redaction. In this paper, we introduce \ciphr, for \textbu{C}ryptography-inspired \textbu{I}\textbu{P} protection through fine-grain \textbu{H}ardware \textbu{R}edaction, to address the limitations of prior art. \ciphrs borrows the concept of indistinguishability from cryptography and employs randomized structural modifications using configurable blocks that are programmable using functional bitstream to redact security-critical logic from the design. \ciphrs significantly improves the brute-force RE attack complexity and improves structural attack resilience, which is demonstrated using multiple open-source benchmarks.

The paper makes the following major contributions.
\begin{enumerate}
    \item It identifies a critical deficiency in existing hardware IP protection technologies -- namely, their inability to protect against known design attacks or library attacks -- a threat model, which enables an adversary to extract design secrets from a protected design by identifying its similarity to a library of known designs.   
    \item Based on this observation, it mathematically associates the IP protection problem against the above threat model as the one that achieves the indistinguishability property for a hardware design -- analogous to the ciphertext indistinguishability property of many encryption techniques. 
    \item It presents design transformations based on hardware redaction that achieves functional and structural indistinguishability as demonstrated through quantitative metrics and mathematical analysis. It also presents a complete tool flow to achieve the proposed transformation at acceptable hardware overhead (comparable to the existing fine-grain redaction approaches) in terms of area and performance.
    \item Using open-source design benchmarks from the ISCAS89, ITC99, and MIT-CEP suites, the paper presents extensive evaluation results\footnote{\label{note1}\textcolor{blue}{Access to the \ciphrs artifact repository available upon request.}} on the effectiveness of the proposed transformation to protect against the above attack vector.
\end{enumerate}

The paper is organized as follows: Section \ref{sec:back} discusses the zero-trust model and explores existing IP-level countermeasures. Section \ref{sec:motiv} outlines the threat model and motivation behind this work. Section \ref{sec:method} describes the major steps of the proposed methodology. Section \ref{sec:tdi} introduces the metrics used for quantifying transformed design indistinguishability. Section \ref{sec:result} presents implementation results and security evaluation of the proposed methodology followed by a comparative analysis. Section \ref{sec:conclude} concludes the paper.

\section{Background}
\label{sec:back}
In this section, we describe the threats associated with the zero-trust model and explore different countermeasures for protecting IP confidentiality and integrity against RE attacks. 

\subsection{Hardware Zero-Trust Model}

The IC supply chain involves numerous steps, from design and fabrication to packaging and distribution, often involving multiple stakeholders and third-party vendors. The current globalization and increasing demand to meet time-to-market have led the semiconductor industry to increasingly outsource critical steps in the IC manufacturing process to foreign entities. While this approach offers cost advantages and faster production cycles, it also introduces significant security risks. These external entities may inadvertently or maliciously compromise the integrity of the IC design, introduce counterfeit components, overproduction, or insert hardware Trojans \cite{hw_security_book_2018_Bhunia} that can significantly
undermine the chip’s reliability and trustworthiness. By assuming the zero-trust model, where any entity in the supply chain could be compromised, it is essential to ensure the protection of sensitive assets of a design and provide a higher level of security against a range of threats.

\subsection{Anti-RE IP Protection Landscape}

\begin{figure}[!htbp]
    \centering
    \includegraphics[width=0.9\columnwidth]{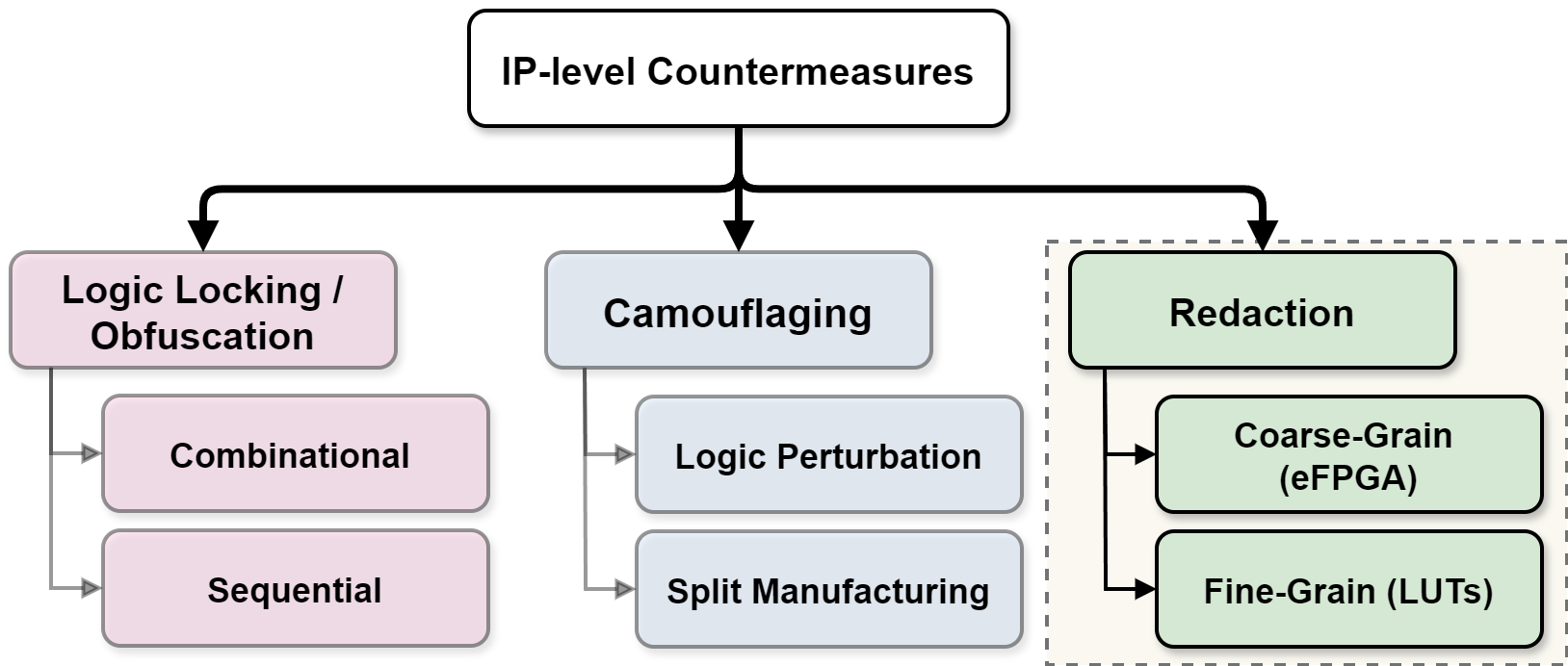}
    \caption{Taxonomy of various IP protection schemes against piracy and RE.}
    \label{fig:taxonomy}
\end{figure}

Reverse engineering poses a significant threat to the IC supply chain by enabling attackers to extract sensitive design information, leading to potential misuse or unauthorized access. To counter RE attacks, various countermeasures have been proposed, broadly classified into obfuscation, camouflaging, and redaction, as shown in Fig. \ref{fig:taxonomy}.

\begin{itemize}
    \item \textbf{Logic Locking/Obfuscation} conceals a design’s original functionality using key-based validation, making reverse engineering more challenging. It is categorized into combinational and sequential obfuscation techniques.
    \item \textbf{Camouflaging} introduces significant design modifications, such as logic perturbation and split manufacturing, to mislead attackers without altering the functionality of the system. Split manufacturing \cite{split_manu}, for instance, logically separates different parts of the design, which can complicate reassembly and analysis, especially when an adversary lacks knowledge of the varying processes or technologies used in each segment.
    \item \textbf{Redaction} selectively hides critical components to prevent RE. It is divided into fine-grain (LUT-based) and coarse-grain (eFPGA Macro-based) approaches, as illustrated in Fig. \ref{fig:rd_types}. Both methods rely on protected bitstreams to configure reconfigurable fabrics, ensuring that even if the design is manufactured in an untrusted foundry, adversaries cannot access critical design details.
\end{itemize}

\section{Motivation}
\label{sec:motiv}

In this section, we discuss the assumed threat model and introduce the cryptographic concepts of Kerckhoffs’s principle and indistinguishability, followed by the limitations of current redaction techniques and the motivation behind this work. 

\subsection{Threat Model}

Existing solutions for hardware IP redaction, both LUT-based \cite{LUT-Lock_2018, Custom-LUT_2019,EvoLUTe_2023} and eFPGA-based \cite{eFPGA-CMU-2_2021, eFPGA-NYU_2021, ALICE_2022, SheLL_2023,fpga_obf_survey} techniques, focus solely on their resistance to functional \textit{oracle-guided} attacks \cite{SAT_2015,smt_attack,icysat} and completely ignore the possibility of \textit{oracle-less} RE attacks \cite{SAIL_2018, SWEEP_2019,DANA_2020}. Moreover, traditional threat models for hardware IP protection do not consider the risks posed by a skilled adversary with privileged access to the IC supply chain who can leverage prior knowledge about security countermeasures and utilize commercial EDA tools to execute more sophisticated \textit{oracle-less} attacks \cite{gnnunlock,dasgupta2025latte,knowledge_guided_attack}.  Furthermore, in an untrusted foundry environment, the programming bitstream can be vulnerable to \textit{physical attacks} such as side-channel analysis \cite{SCA_2024_Tajik} and microprobing \cite{Probing_2017_Tajik}. Since the proposed methodology in \ciphrs builds upon conventional hardware redaction techniques \cite{hipr_2025_tches} that are proven to be secure against \textit{oracle-guided} attacks, we limit our security evaluation to the mitigation of emerging \textit{oracle-less} RE attacks \cite{dasgupta2025latte, knowledge_guided_attack} carried out by a privileged adversary. Similarly, existing \textit{oracle-guided} attacks \cite{Break-and-Unroll_2022, FuncTeller_2023} demonstrated on eFPGA-based redaction techniques that leverage SAT-solvers are not explored in this work. We assume that existing countermeasures for bitstream protection in programmable fabrics such as FPGAs can be replicated in \ciphr, and are deemed beyond the scope of this work. 

For a comprehensive security analysis of \ciphr, we adopt the following zero-trust threat model with an emphasis on \textit{oracle-less} structural attacks:

$\circ$ \textbf{Assets:} RTL or Gate-level hardware IPs (with associated design files).

$\circ$ \textbf{Adversary:} SoC design house (integration, synthesis, and physical layout); Untrusted third-party facilities (fabrication, package \& test); End users (post-deployment).

$\circ$ \textbf{Adversarial Access:} Skilled entities with privileged access to the redacted IP, prior knowledge about the redaction technique, commercial EDA tools to perform structural analysis-based \textit{oracle-less} attacks on redacted netlist.

$\circ$ \textbf{Adversarial Objectives:} Recover the correct bitstream that enables the original IP functionality and/or extract security-critical design secrets.

$\circ$ \textbf{Trust Model:} IP vendor/owner/designer is trustworthy.

\subsection{Kerckhoffs's Principle \& Indistinguishability}

Kerckhoffs's Principle \cite{book_crypto_katz} states that a cryptographic system should be secure even if everything about the system, except the key, is public knowledge and is applicable to all contemporary encryption algorithms (AES, RSA, etc.) that are considered to be secure and thoroughly investigated. The security of the encrypted message depends solely on the security of the secret encryption key. IP-level countermeasure techniques adhering to Kerckhoffs's principle are crucial to ensure robust security against RE and cloning attacks.

Indistinguishability \cite{book_crypto_katz, IND_OBF_jain_2021} refers to the property of a cryptosystem in which two or more items (such as plaintexts, ciphertexts, or keys) cannot be distinguished from each other by an adversary with a probability significantly higher than random guessing ($1/2$), even if the adversary has prior knowledge of the cryptosystem. Under a known-plaintext attack (KPA) \cite{book_crypto_katz}, a cryptosystem is \textit{indistinguishable} if the adversary can correctly distinguish the chosen ciphertext with the probability:
\setlength{\belowdisplayskip}{0pt} \setlength{\belowdisplayshortskip}{0pt}
\setlength{\abovedisplayskip}{0pt} \setlength{\abovedisplayshortskip}{0pt}
\begin{dmath}\label{eq:p_ind}
    {\mathcal{P}_{ind}} = 1/2 + \xi(\lambda)
\end{dmath}
where $\lambda$ is the security parameter (e.g., key size) and $\xi$ is a \textit{negligible function} \cite{IND_OBF_jain_2021}, implying that $\mathcal{P}_{ind}$ corresponds to a \textit{negligible} advantage for the adversary over random guessing.

\begin{figure}[!htbp]
    \centering
    \subfloat[Types of Hardware IP Redaction.]
    {\includegraphics[width=1.0\columnwidth]{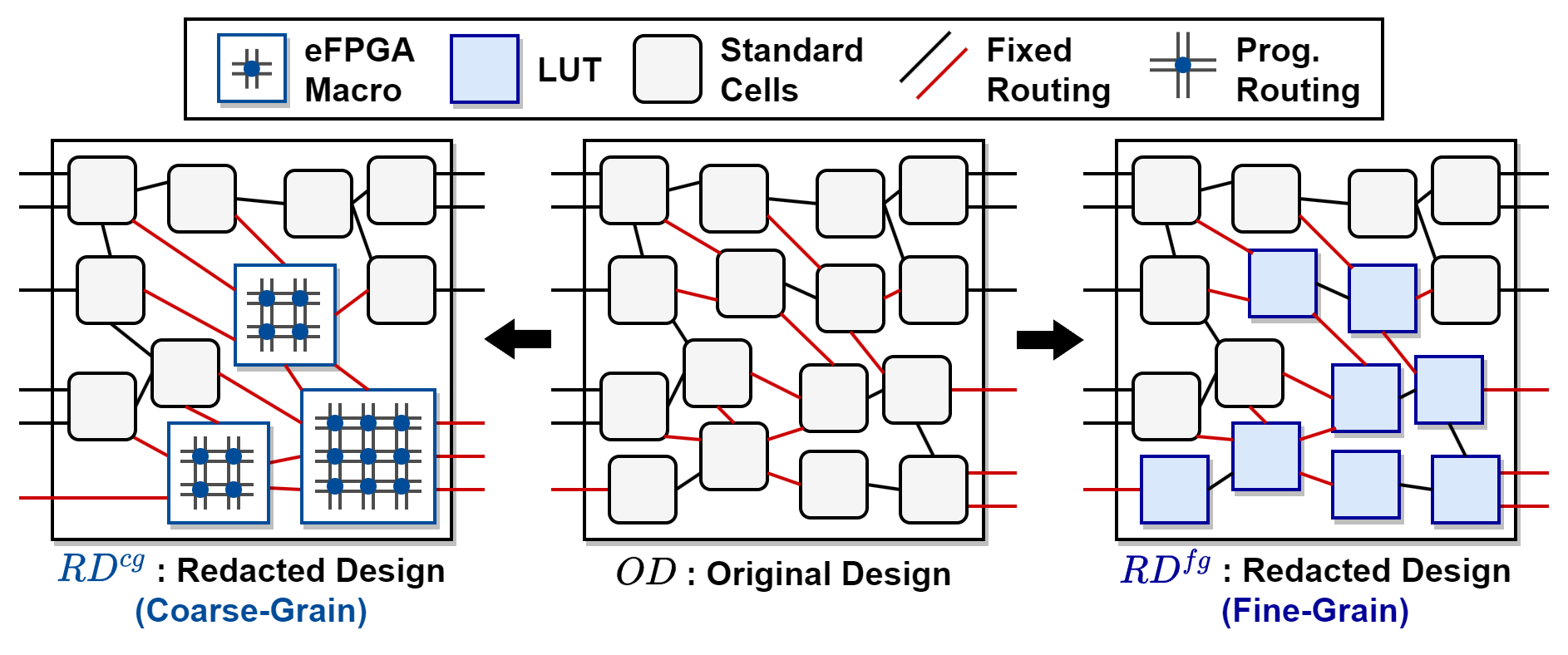}
    \label{fig:rd_types}}
    \hfill
    \subfloat[\textit{LATTE} can identify $\{OD_{i}, \textcolor{RoyalBlue}{RD_{i}^{cg}}, \textcolor{Blue}{RD_{i}^{fg}\}}$.]
    {\includegraphics[width=1.0\columnwidth]{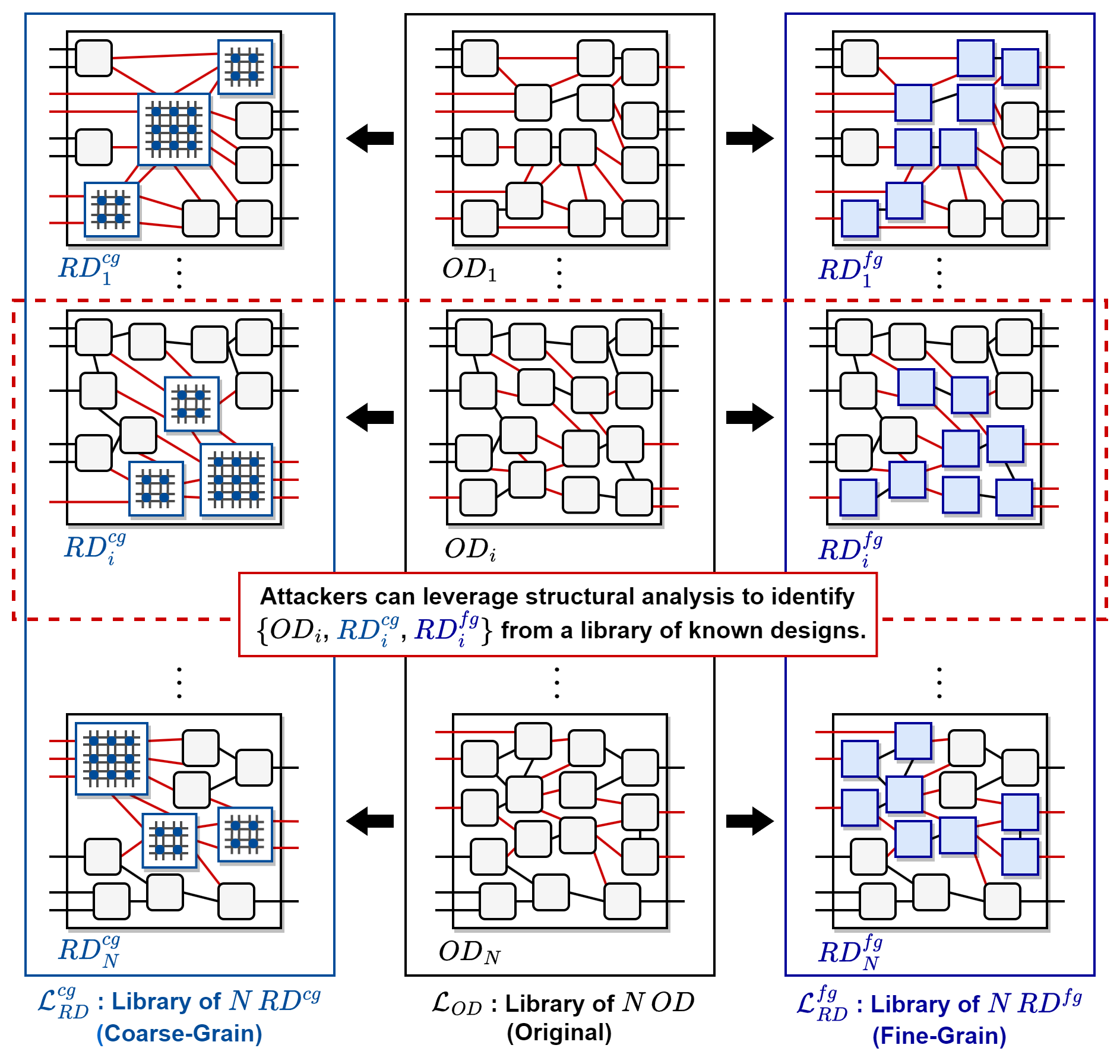}
    \label{fig:rd_libAtk}}
    \caption{Overview of existing redaction techniques: (a) Coarse-grain (eFPGA-based) vs fine-grain (LUT-based) redaction; The hardware components and routing are not to scale. (b) The fixed interconnects (denoted in \textcolor{Maroon}{red}) in the $OD$ are not randomized during either type of redaction, leading to structural rigidness in both $RD^{cg}$ and $RD^{fg}$ that make them vulnerable to \textit{oracle-less} structural attacks like \textit{LATTE}~\cite{dasgupta2025latte,library_attack_2025_arXiv}.}
    \label{fig:rd_back}
\end{figure}

\subsection{Motivation for \ciphr}

Current state-of-the-art fine-grain and coarse-grain redaction techniques introduce limited structural and functional variation, leaving them vulnerable to sophisticated \textit{oracle-less} attacks that can lead to recovery of the original unprotected design. Fig. \ref{fig:rd_types} shows how existing redaction techniques in the literature fail to modify the fixed interconnections between redacted IP blocks. This structural rigidness makes the redacted logic vulnerable against structural attacks like \textit{LATTE}~\cite{dasgupta2025latte,library_attack_2025_arXiv} under the threat model assumed for security evaluation. \textit{LATTE} represents the real-life scenario where a highly skilled and privileged adversary in the IC supply chain \cite{intel_threat_model} can leverage a library of known IPs and the IP protection countermeasure to recover the original design, emulating KPA on a cryptosystem. To address this limitation, \ciphrs incorporates novel randomizations to significantly transform the configurable fabric and its components during redaction, making it resistant to such \textit{oracle-less} attacks based on structural analysis. By extending the cryptographic principle of indistinguishability to hardware IP protection, \ciphrs ensures that an adversary cannot differentiate between functionally equivalent but structurally distinct implementations, thus obscuring critical design information and significantly enhancing security against RE. Table \ref{tab:background} presents a comparative analysis between \ciphrs and existing redaction techniques in literature, highlighting the necessity of \ciphrs for robust and effective hardware IP protection.

\begin{table}
\centering
\caption{Comparing \ciphrs with existing redaction techniques.}
\label{tab:background}
\resizebox{1.0\columnwidth}{!}{%
\begin{tabular}{@{}c@{\ho}|@{}c@{\ho}|@{\ho}c@{\ho}|@{\ho}c@{\ho}|@{\ho}c@{\ho}|@{\ho}c@{\ho}|@{\ho}c@{}}
\hline
\multirow{2}{*}{\textbf{Technique}} & 
\multirow{2}{*}{\ho\textbf{Method}} & 
{\ho\textbf{Input}} & 
{\ho\textbf{Tool}} & 
\multirow{2}{*}{\ho\textbf{$\mathbb{RT}$?}} & 
\multirow{2}{*}{\ho\textbf{$\mathbf{TDI}$?}} & 
{\ho\textbf{\textit{LATTE}} \cite{dasgupta2025latte}} \\
{} &
{} & 
{\ho\textbf{Abstract$^{\mathbf{n}}$}} &  
\ho\textbf{Flow\ddg?} & 
{} & 
{} & 
\textbf{Resistant?} \\
\hline
\textbf{eFPGA$_{\mathbf{1}}$} \cite{eFPGA-CMU-2_2021}   & eFPGA    & RTL     & \redx & \redx & \redx & \redx \\
\textbf{eFPGA$_{\mathbf{2}}$} \cite{eFPGA-NYU_2021}     & eFPGA    & RTL     & \redx & \redx & \redx & \redx \\
\textbf{ALICE} \cite{ALICE_2022}                        & eFPGA    & RTL     & \grnc & \redx & \redx & \redx \\
\textbf{SheLL} \cite{SheLL_2023}                        & eFPGA    & RTL     & \grnc & \redx & \redx & \redx \\
\hdashline
\textbf{LUT-Lock} \cite{LUT-Lock_2018}                  & LUT      & GL      & \redx & \redx & \redx & \redx \\
\textbf{Custom-LUT} \cite{Custom-LUT_2019}              & LUT      & GL      & \redx & \redx & \redx & \redx \\
\textbf{EvoLUTe} \cite{EvoLUTe_2023}                    & LUT      & RTL, GL & \grnc & \redx & \redx & \redx \\
\textbf{HIPR} \cite{hipr_2025_tches}                    & LUT      & RTL, GL & \grnc & \redx & \redx & \grnc \\
\ciphr\bstars                                           & LUT      & RTL, GL & \grnc & \grnc & \grnc & \grnc \\
\hline
\end{tabular}%
}
\footnotesize
\raggedright\\
\bstars \textbf{Current work.} \textbf{GL: Gate-Level.} \textbf{$\mathbb{RT}$: Randomized Transformations.} \\
\ddg \textbf{Automated Tool Flow.} \textbf{$\mathbf{TDI}$: Transformed Design Indistinguishability.} 
\end{table}

\section{\ciphrs Methodology}
\label{sec:method}

\begin{figure}[!htbp]
    \centering
    \includegraphics[width=0.9\columnwidth]{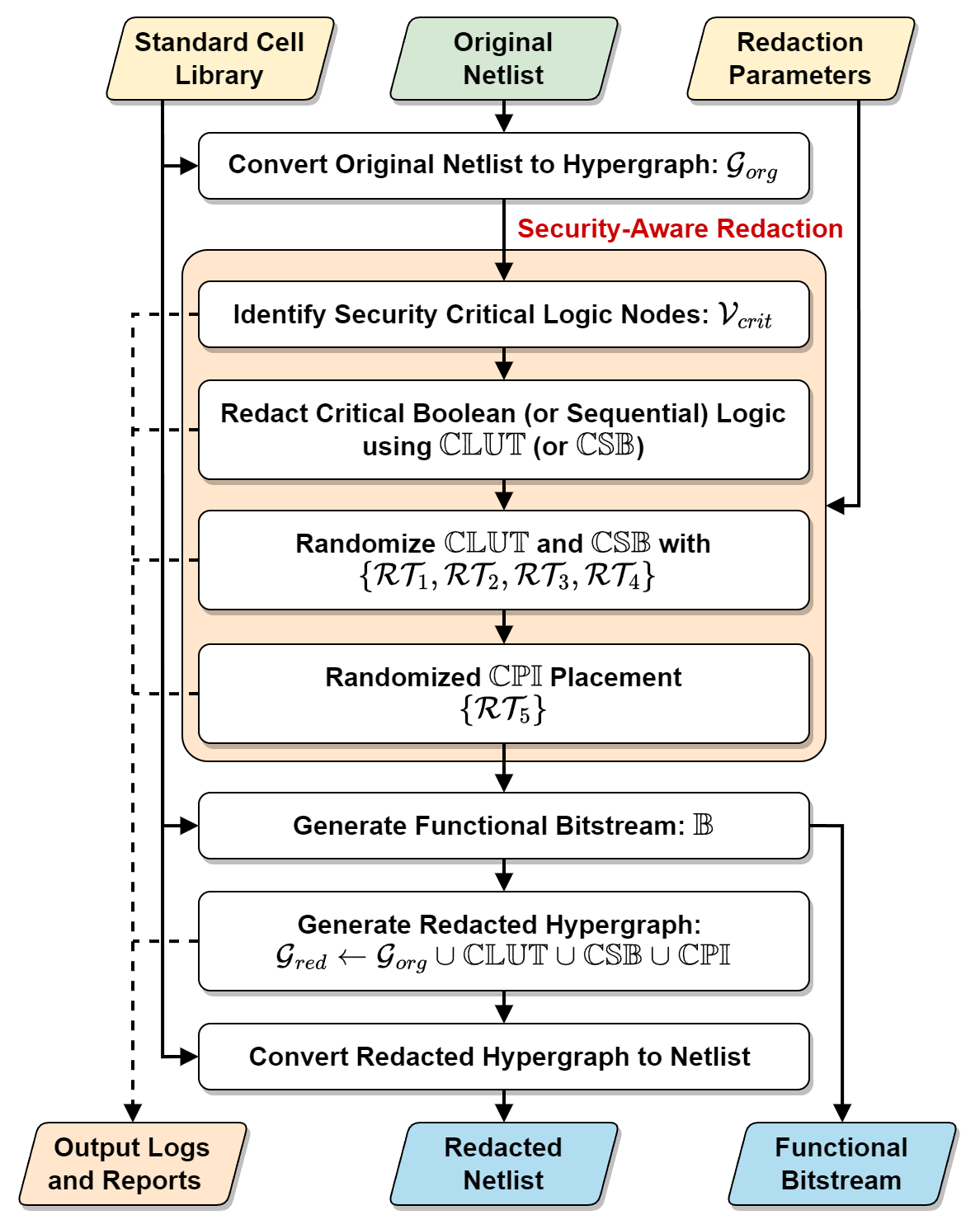}
    \caption{Major steps in the proposed \ciphrs tool flow. \ciphrs applies novel randomized transformations ($\mathbb{RT}$) over existing IP protection techniques such as fine-grain redaction \cite{hipr_2025_tches} to achieve functional and structural indistinguishability.}
    \label{fig:ciphr_flow}
\end{figure}

The major steps of the proposed \ciphrs methodology are presented in Fig. \ref{fig:ciphr_flow}. First, the original gate-level netlist $N_{org}$ is converted into a hypergraph $\mathcal{G}_{org} = (V,E)$ with logic gates and flip-flops (FFs) represented as vertices ($V$), with the interconnects becoming edges ($E$). In \ciphr, we replicate the fine-grain redaction steps from prior work \cite{hipr_2025_tches} and enhance them by introducing novel randomized transformations, which can be customized using the redaction parameter $\gamma$ and the random seed $\theta$. $\mathcal{G}_{org}$ is topologically sorted and the set of critical nodes $\mathcal{V}_{crit}$ is identified for redaction using the removal cost function ($rcf$) \cite{hipr_2025_tches} that depends upon structural characteristics such as fan-in/fan-out cone sizes as well as stochastic properties such as Shannon entropy \cite{hw_security_book_2018_Bhunia}. The redacted logic is implemented using a configurable fabric $\{\mathbb{CLUT},\mathbb{CSB},\mathbb{CPI}\}$ that can be programmed using the corresponding bitstream $\mathbb{B}$ to restore its true functionality. The configurable fabric is merged with the remaining original design logic to obtain the redacted hypergraph $\mathcal{G}_{red}$, which is then converted into the redacted netlist $N_{red}$. Algorithm \ref{algo:ciphr} outlines the various stages involved in \ciphrs methodology. 
\setlength{\textfloatsep}{0pt}
\begin{algorithm}
\footnotesize
\caption{\ciphrs Main}
\label{algo:ciphr}
    \SetKwInOut{Input}{Input}
    \SetKwInOut{Output}{Output}
    \textbf{Procedure} \textbf{\textit{ciphr\_main}}\\
    \Input{Original Netlist ($N_{org}$), Random Seed ($\theta$), Redaction Parameters ($\gamma$)}
    \Output{Redacted Netlist ($N_{red}$), Functional Bitstream ($\mathbb{B}$)}

    $\mathcal{G}_{org} \leftarrow$ \textbf{\textit{netlist\_to\_hypergraph}}($N_{org}$) \\
    $\mathcal{G}_{sort} \leftarrow$ \textbf{\textit{topological\_sort}}($G_{org}$) \\
    
    \tcc{Security-Aware Fine-Grain Redaction}
    $\mathcal{V}_{crit} \leftarrow$ \textbf{\textit{identify\_critical\_nodes}}($\mathcal{G}_{sort},\theta,\gamma$) \\
    $\{\mathbb{CLUT},\mathbb{CSB},\mathbb{CPI}\} \leftarrow$ \textbf{\textit{redact\_critical\_logic}}($\mathcal{G}_{sort}, \mathcal{V}_{crit}, \theta, \gamma$) \\
    $\mathcal{G}_{red} \leftarrow \mathcal{G}_{sort} \cup \mathbb{CLUT} \cup \mathbb{CSB} \cup \mathbb{CPI}$ \\
    
    \tcc{Generate Functional Bitstream}
    $\mathbb{B} \leftarrow$ \textbf{\textit{generate\_bitstream}}($\mathcal{G}_{red}, \theta$) \\

    $N_{red} \leftarrow$ \textbf{\textit{hypergraph\_to\_netlist}}($\mathcal{G}_{red}$) \\
    
    \textbf{return} $N_{red}, \mathbb{B}$ 
    
\end{algorithm}

\subsection{Configurable Fabric}

\begin{figure}[!htbp]
    \centering
    \subfloat[]{\includegraphics[width=0.7\columnwidth]{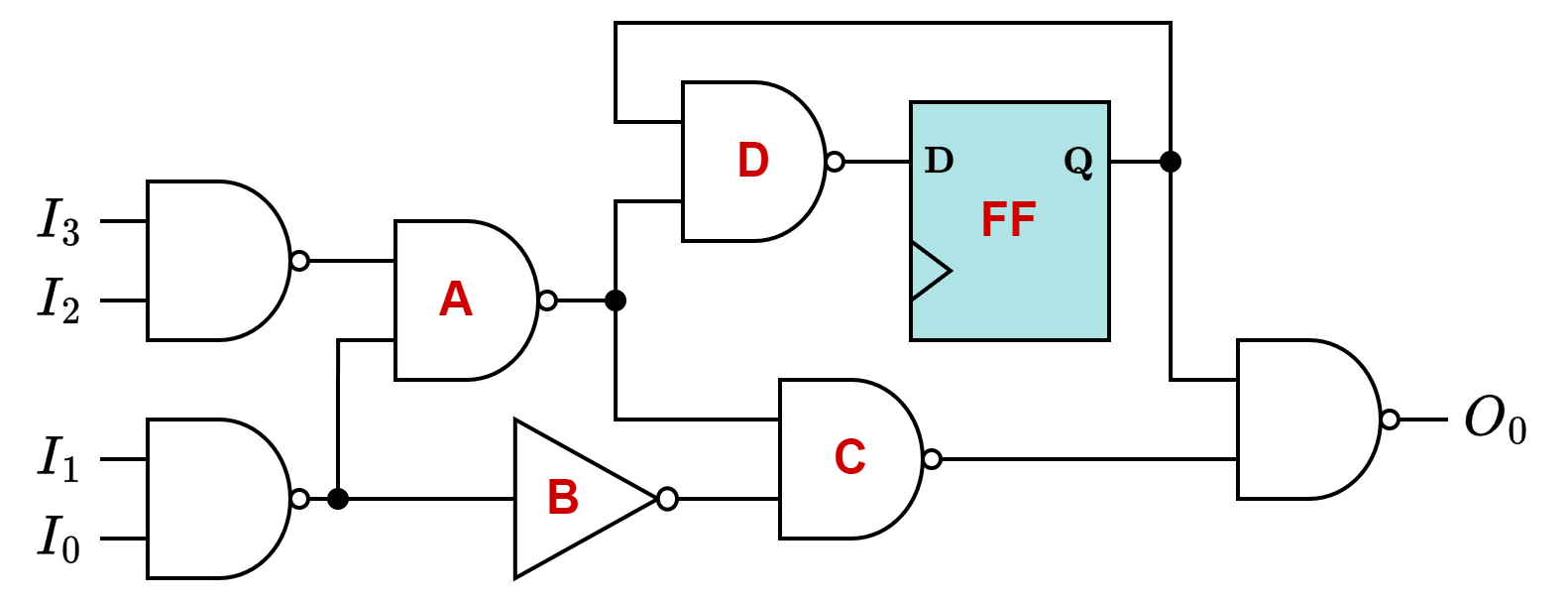}}
    \hfill
    \subfloat[]{\includegraphics[width=0.7\columnwidth]{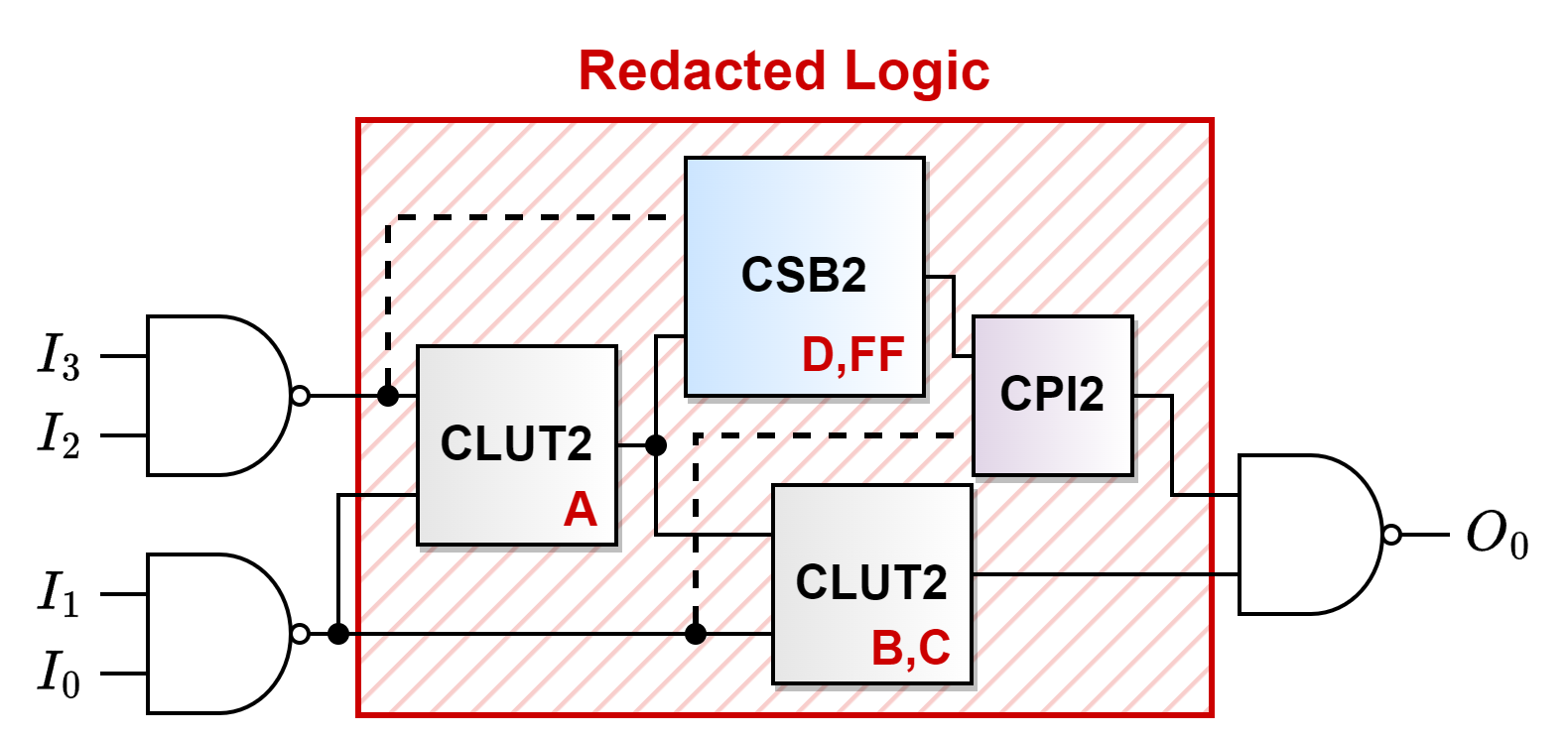}}
    \caption{Overview of fine-grain hardware redaction in \ciphr: (a) original gate-level netlist, (b) redacted netlist with a configurable fabric implementing the redacted Boolean, sequential and interconnect logic.}
    \label{fig:hw_redaction}
\end{figure}

Fig. \ref{fig:hw_redaction} shows how the security-critical logic in a gate-level netlist is redacted using the configurable fabric in \ciphr, which consists of the following programmable components:

\begin{itemize}
    \item \textbf{Configurable Look-Up Tables (CLUTs)}: redacts Boolean logic gates.
    \item \textbf{Configurable Sequential Blocks (CSBs}): redacts sequential logic (FFs). Each CSB also contains a CLUT for either implementing existing Boolean logic or introducing dummy logic before the redacted FF.
    \item \textbf{Configurable Programmable Interconnects (CPIs}): redacts interconnect information and randomizes them. 
\end{itemize} 
For the remainder of this paper, CLUT$n$/CSB$n$/CPI$n$ corresponds to a CLUT/CSB/CPI with $n$ inputs and 1 output.

\begin{figure}[!htbp]
    \centering
    \includegraphics[width=0.7\columnwidth]{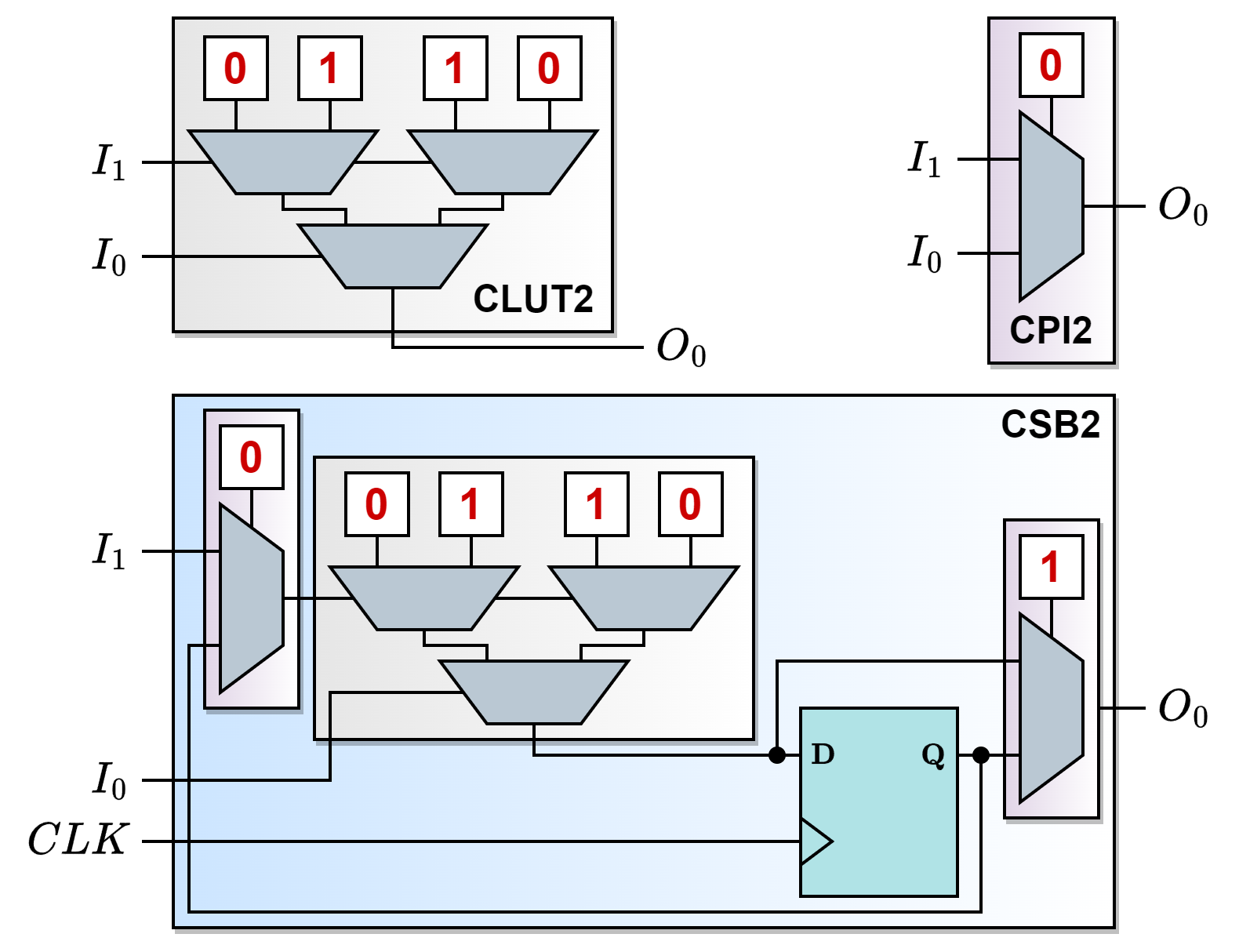}
    \caption{The programmable components of the configurable fabric used for fine-grain redaction in \ciphr. The gate-level implementations of CLUT2, CPI2 and CSB2 are realized using typical standard cells: 2x1 MUXes, bitstream registers and data flip-flops.}
    \label{fig:cfg_fabric}
\end{figure}

Fig. \ref{fig:cfg_fabric} illustrates how the components of configurable fabric in \ciphrs are realized at the gate level using 2x1 MUXes, bitstream registers, and data FFs from a standard cell library. Redacted logic is replaced with CLUT/CSB/CPI instances in RTL to avoid dependencies and mapped to the target library via constrained synthesis. Bitstream registers are daisy-chained to form one or more shift registers for serial loading. For optimal PPA results, the CLUT/CSB/CPI(s) should be custom-designed to leverage transistor-level optimizations along with better placement and routing.

\setlength{\textfloatsep}{0pt} 
\begin{algorithm}
\footnotesize
\caption{Security-Aware Fine-Grain Redaction}
\label{algo:redact}
    \SetKwInOut{Input}{Input}
    \SetKwInOut{Output}{Output}
    \textbf{Procedure} \textbf{\textit{redact\_critical\_logic}}\\
    \Input{$\mathcal{G}_{sort}, \mathcal{V}_{crit}, \theta, \gamma$}
    \Output{List of CLUTs ($\mathbb{CLUT}$), CSBs ($\mathbb{CSB}$), CPIs ($\mathbb{CPI}$)}

    $\mathbb{CLUT} \leftarrow \emptyset$, $\mathbb{CSB} \leftarrow \emptyset$, $\mathbb{CPI} \leftarrow \emptyset$ \\

    \ForEach{$\mathcal{V} \in \mathcal{V}_{crit}$}   
    {
        $\mathcal{V}_{MFFC} \leftarrow$ \textbf{\textit{extract\_MFFC}}($\mathcal{G}_{sort}, \mathcal{V}$) \\ 
        $r\_size \leftarrow \gamma_{min} + \textbf{\textit{random}}(\theta) \% (\gamma_{max} - \gamma_{min} + 1)$ \\
            
        \If{\textbf{\textit{is\_FF}}($\mathcal{V}$)}
        {
            \tcc{Redact Sequential Logic into CSB}
            $CSB_{i} \leftarrow$ \textbf{\textit{redact\_sequential\_logic}}($\mathcal{V}_{MFFC}, r\_size, \gamma$) \\
            $\mathbb{CSB}$.\textbf{\textit{append}}($CSB_{i}$) \\
        }
        \Else
        {
            \tcc{Redact Boolean Logic into CLUT}
            $CLUT_{i} \leftarrow$ \textbf{\textit{redact\_boolean\_logic}}($\mathcal{V}_{MFFC}, r\_size, \gamma$) \\
            $\mathbb{CLUT}$.\textbf{\textit{append}}($CLUT_{i}$) \\
        }
        
    }

    \tcc{Randomized (Dummy) CSB Placement}
    $n_a \leftarrow$ \textbf{\textit{random}}($\theta$) \% ($\gamma_{a\_max} \times |\mathbb{CLUT}|$) \\
    \While{$n_a > 0$}
    {
        $CLUT_{i} \leftarrow$ \textbf{\textit{find\_rand\_clut}}($\mathbb{CLUT}, \theta$) \\
        \If{\textbf{\textit{is\_suitable\_csb}}($CLUT_{i}$)}
        {
            $CSB_{i} \leftarrow$ \textbf{\textit{convert\_clut\_to\_csb}}($CLUT_{i}, \mathcal{G}_{sort}, \gamma$)\\
            $\mathbb{CSB}$.\textbf{\textit{append}}($CSB_{i}$); $\mathbb{CLUT}$.\textbf{\textit{remove}}($CLUT_{i}$)  \\
        }
        $n_a \leftarrow n_a - 1$
    }
    $n_b \leftarrow$ \textbf{\textit{random}}($\theta$) \% ($\gamma_{b\_max} \times |\mathcal{G}_{sort}.\mathcal{V}_{FF}|$) \\
    \While{$n_b > 0$}
    {
        $FF_{i} \leftarrow$ \textbf{\textit{find\_rand\_flop}}($\mathcal{G}_{sort}.\mathcal{V}_{FF}, \theta, \gamma$) \\
        \If{\textbf{\textit{is\_suitable\_csb}}($FF_{i}$)}
        {
            $CSB_{i} \leftarrow$ \textbf{\textit{convert\_flop\_to\_csb}}($FF_{i}, \mathcal{G}_{sort}, \gamma$)\\
            $\mathbb{CSB}$.\textbf{\textit{append}}($CSB_{i}$) \\
        }
        $n_b \leftarrow n_b - 1$
    }

    \tcc{Randomize CLUTs and CSBs}
    \ForEach{$CLUT_{i} \in \mathbb{CLUT}$}
    {
        \If{\textbf{\textit{is\_even}}(\textbf{\textit{random}}($\theta$))}
        {
            $CLUT_{i} \leftarrow$ \textbf{\textit{add\_dummy\_inputs}}($CLUT_{i}, \mathcal{G}_{sort}, \theta, \gamma$) \\
            $CLUT_{i} \leftarrow$ \textbf{\textit{reorder\_inputs}}($CLUT_{i}, \mathcal{G}_{sort}, \theta, \gamma$) \\
            $CLUT_{i} \leftarrow$ \textbf{\textit{invert\_output}}($CLUT_{i}, \mathcal{G}_{sort}, \theta, \gamma$) \\
        }
    }

    \ForEach{$CSB_{i} \in \mathbb{CSB}$}
    {
        \If{\textbf{\textit{is\_even}}(\textbf{\textit{random}}($\theta$))}
        {
            $CSB_{i} \leftarrow$ \textbf{\textit{add\_dummy\_inputs}}($CSB_{i}, \mathcal{G}_{sort}, \theta, \gamma$) \\
            $CSB_{i} \leftarrow$ \textbf{\textit{reorder\_inputs}}($CSB_{i}, \mathcal{G}_{sort}, \theta, \gamma$) \\
            $CSB_{i} \leftarrow$ \textbf{\textit{invert\_output}}($CSB_{i}, \mathcal{G}_{sort}, \theta, \gamma$) \\
        }
    }
    
    \tcc{Randomized CPI Placement}
    $\mathcal{E}_{cpi} \leftarrow$ \textbf{\textit{identify\_candidate\_wires}}($\mathcal{G}_{sort}, \theta, \gamma$) \\
    \ForEach{$\mathcal{E} \in \mathcal{E}_{cpi}$}
    {
        \If{\textbf{\textit{is\_suitable\_cpi}}($\mathcal{E}$)}
        {
            $CPI_{i} \leftarrow$ \textbf{\textit{redact\_interconnect\_logic}}($\mathcal{E}, \mathcal{G}_{sort}, \gamma$) \\
            $\mathbb{CPI}$.\textbf{\textit{append}}($CPI_{i}$) \\
        }
    }
    
    \textbf{return} $\{\mathbb{CLUT},\mathbb{CSB},\mathbb{CPI}\}$ 
    
\end{algorithm}

\subsection{Security-Aware Fine-Grain Redaction}

Algorithm \ref{algo:redact} describes the security-aware fine-grain redaction proposed in \ciphr. A CLUT/CSB with randomly chosen size ($r\_{size}$) is used to redact the appropriate Boolean and/or sequential logic from the maximum fan-out free cone (MFFC) \cite{ABC_ICCAD2007_Mishchenko} corresponding to each critical vertex $\mathcal{V} \in \mathcal{V}_{crit}$. The \ciphrs tool flow incorporates five distinct types of novel and randomized transformations encapsulated in $\mathbb{RT} = \{\mathcal{RT}_{1},\mathcal{RT}_{2},\mathcal{RT}_{3},\mathcal{RT}_{4},\mathcal{RT}_{5}\}$ in the configurable fabric to achieve functional and structural indistinguishability during fine-grain redaction. The transformations in $\mathbb{RT}$ are randomized with the 4-byte integer seed $\theta$ and can be customized using the redaction parameter $\gamma$. 

\subsubsection{Randomized CLUT/CSB Mapping}

\ciphrs applies randomized CLUT/CSB mapping ($\mathcal{RT}_{1}$) to prevent the deterministic mapping of security-critical Boolean and sequential logic to CLUTs and CSBs, respectively, increasing the RE complexity of the redacted netlist. $\mathcal{RT}_{1}$ is achieved by a combination of Fisher-Yates shuffling \cite{Fisher_Yates-AFP} of the $\mathcal{V}_{crit}$ array and the randomly chosen CLUT/CSB size ($r\_{size}$). 

Consider a Boolean logic cone $G_{n\times1}$ with $n$ inputs and a single output, redacted using a configurable fabric containing $m$ different CLUT/CSB sizes. Given a set of CLUT/CSB sizes $S = \{s_i : {\gamma_{min}} \leq s_i \leq {\gamma_{max}} \mid s_i \in \mathbb{N},\,1 \leq i \leq m \}$, where, $s_1<s_2 <\ldots<s_m$. Let $x_i$ be the number of CLUTs/CSBs of size $s_i$. During fine-grain redaction, $G_{n\times1}$ is decomposed into multiple CLUT/CSB(s) such that the total number of inputs covered is represented as: 
\begin{dmath} \label{eq:RT1}
\sum_{i=1}^{m} x_i s_i \geq n
\end{dmath}

\begin{figure}[!htbp]
    \centering
    \includegraphics[width=0.8\columnwidth]{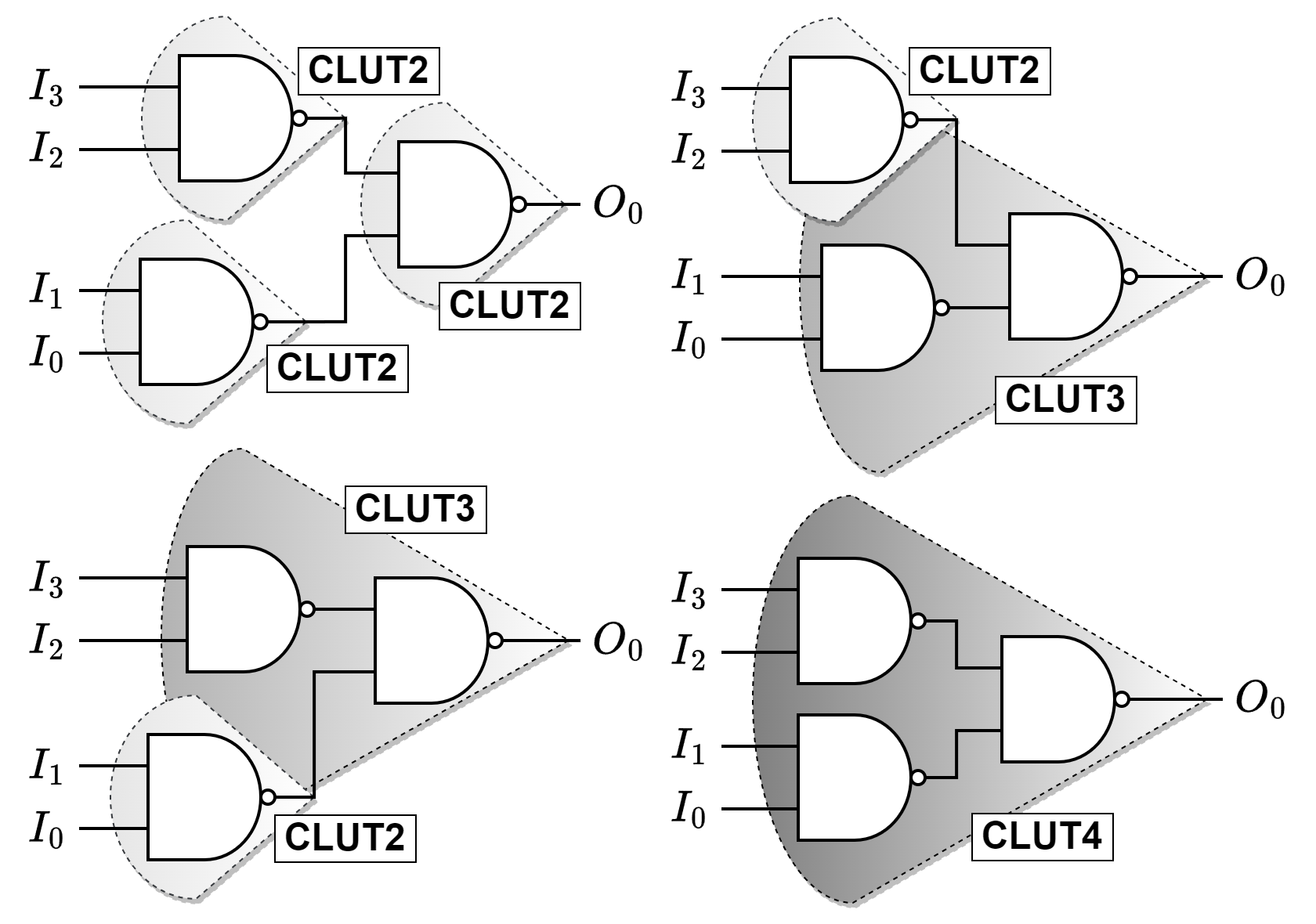}
    \caption{Randomized CLUT/CSB mapping in \ciphrs ($\mathcal{RT}_{1}$). The same Boolean logic can be redacted using 4 distinct variations of the configurable fabric using different combinations of CLUT2/CLUT3/CLUT4.}%
    \label{fig:RT1}%
\end{figure}

Hardware IP blocks from real-world applications can contain more than millions of such Boolean logic cones having a wide range of input size $n$, and applying $\mathcal{RT}_{1}$ leads to an extremely large number of possible variations of the CLUT/CSB mapping in the configurable fabric, as per Eq. \ref{eq:RT1}. Fig. \ref{fig:RT1} depicts how the same Boolean logic ($n=4$) is redacted using multiple distinct variations of configurable fabric using $\mathcal{RT}_{1}$ in \ciphr.

\subsubsection{Randomized CLUT/CSB Input Space Expansion}

\ciphrs incorporates randomized input space expansion in CLUTs/CSBs ($\mathcal{RT}_{2}$), where dummy inputs are added to existing CLUTs/CSBs in the configurable fabric without altering the redacted logic functionality. $\mathcal{RT}_{2}$ improves the brute-force RE attack complexity as the bitstream size increases exponentially with an increase in the input space, and the addition of dummy interconnects results in significant structural transformations. 

\begin{figure}[!htbp]
    \centering
    \subfloat[]{\includegraphics[width=0.5\columnwidth]{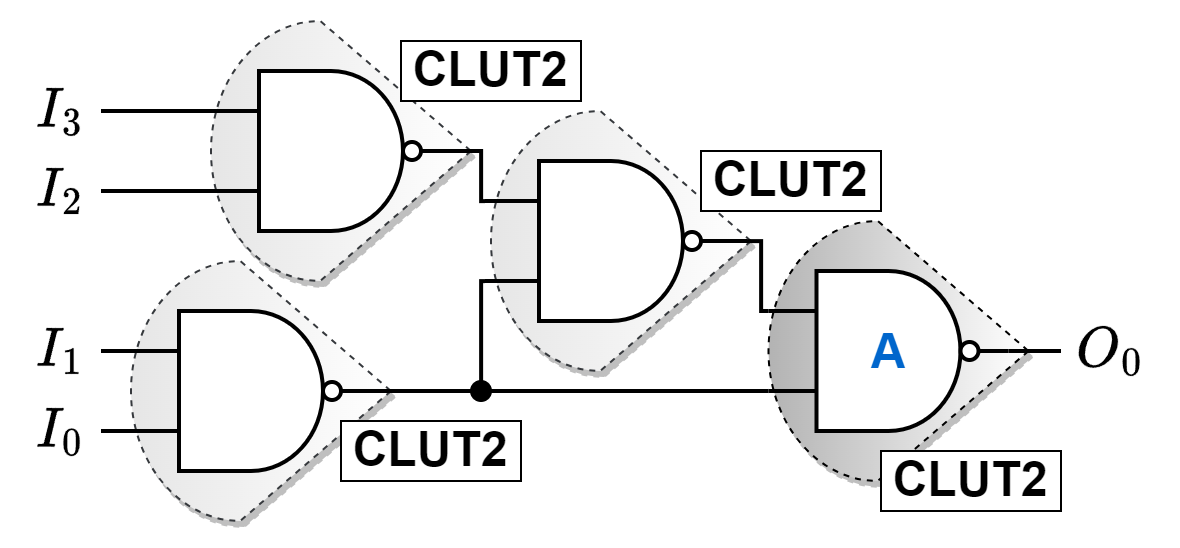}}
    \subfloat[]{\includegraphics[width=0.5\columnwidth]{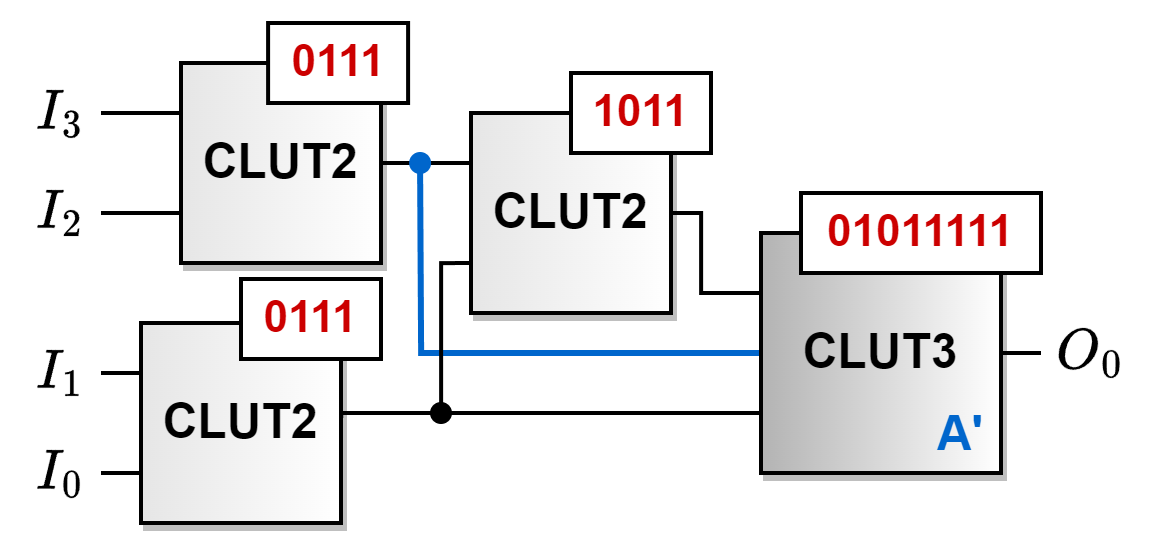}}
    \caption{Randomized CLUT/CSB input space expansion in \ciphrs ($\mathcal{RT}_{2}$): (a) Boolean logic is redacted using only CLUT2, (b) A dummy interconnect (depicted in {\color{RoyalBlue}\textbf{blue}}) is used to convert the CLUT2 for gate {\color{RoyalBlue}\textbf{A}} into a CLUT3, doubling the bitstream size from 4 to 8.}
    \label{fig:RT2}%
\end{figure}

Let $s$ be the initial CLUT/CSB input size, and let $b$ denote the bitstream size such that $b = 2^{s}$. After adding $d$ dummy inputs to the CLUT/CSB, the input size increases to $s' = s+d$ and the new bitstream size becomes $b' = 2^{s'} = 2^{s+d} = 2^d\times2^s$. Hence, the number of Boolean logic functions that can be realized by the CLUT/CSB increases by a factor of $2^d$, increasing brute-force attack complexity. However, the increase in bitstream size doesn’t translate to an exponential increase in security guarantees under \textit{orace-guided} attacks on redaction \cite{eFPGA-IcySAT_2023}. Fig. \ref{fig:RT2} demonstrates how dummy interconnects are added using $\mathcal{RT}_{2}$ to expand the input space for CLUTs in \ciphr. It should be noted that naively adding dummy interconnects can result in undesirable combinational timing loops, and \ciphrs avoids the creation of such loops by selecting dummy interconnects that occur before the current CLUT/CSB in the design topology.

\subsubsection{Randomized CLUT/CSB Functional Space Expansion}

\ciphrs employs randomized functional space expansion in CLUTs/CSBs ($\mathcal{RT}_{3}$) via input reordering and output inversion to maximize the total number of functionalities possible in the configurable without altering the original behavior of the redacted logic. $\mathcal{RT}_{3}$ substantially increases the complexity of RE attacks that employ statistical analysis of the bitstream at a negligible overhead cost.

\begin{figure}[!htbp]
    \centering
    \includegraphics[width=0.7\columnwidth]{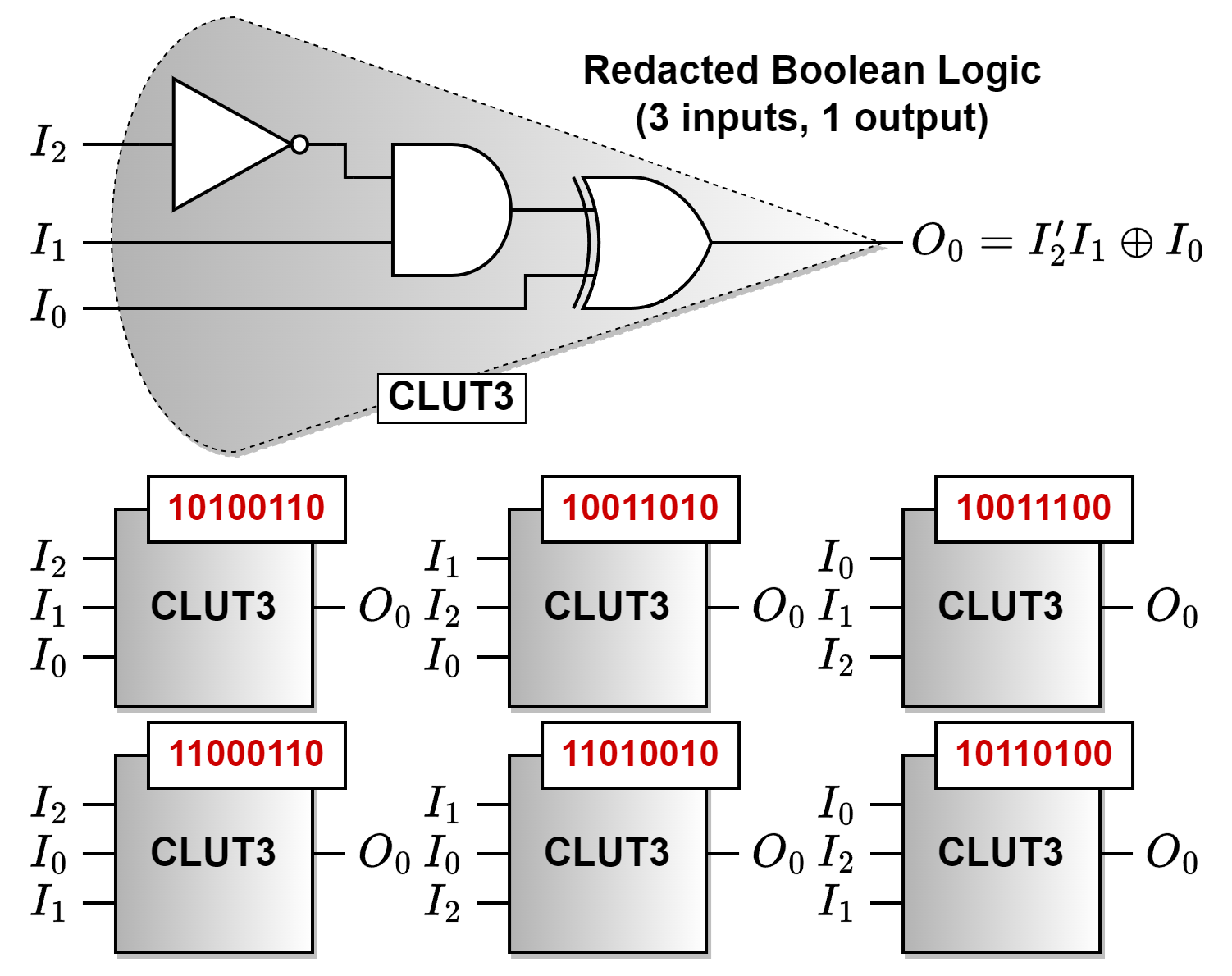}
    \caption{Randomized CLUT/CSB functional space expansion in \ciphrs ($\mathcal{RT}_{3}$). The 3-input Boolean logic function $O_0 = I_2'I_1 \oplus I_0$ is redacted using a CLUT3. The 6 possible permutations of the 3 CLUT inputs ($I_2,I_1,I_0$) and the corresponding bitstream values are demonstrated. Under CLUT output inversion, the bitstream for each permutation will be inverted, resulting in 12 possible bitstream variations for the same redacted logic.}%
    \label{fig:RT3}%
\end{figure}

Let $s$ be the functional input size of the CLUT/CSB, then the total number of input permutations possible is given by $s!$, where each permutation may not result in a unique bitstream value. Since the CLUT/CSB output inversion inverts/flips the bitstream for a given input permutation, the upper bound for unique bitstream values possible due to $\mathcal{RT}_{3}$ is given by $\varphi = 2s!$, for the same logic redacted. Fig. \ref{fig:RT3} shows the randomized input reordering in \ciphrs for a Boolean logic redacted using a CLUT3 and the corresponding bitstream values.

\subsubsection{Randomized CSB Placement}

\begin{figure}[!htbp]
    \centering
    \subfloat[]{\includegraphics[width=0.75\columnwidth]{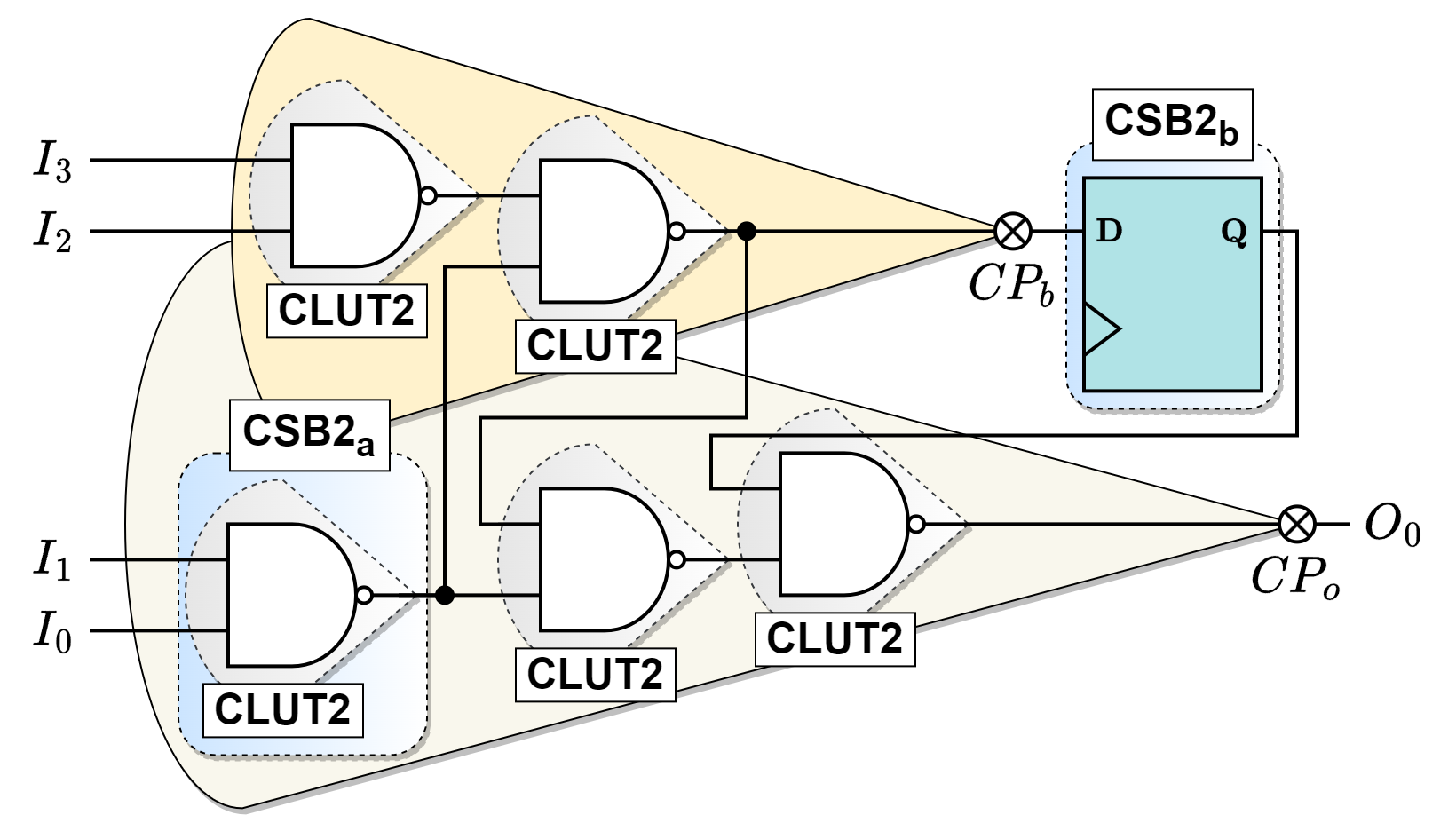}}
    \hfill
    \subfloat[]{\includegraphics[width=0.75\columnwidth]{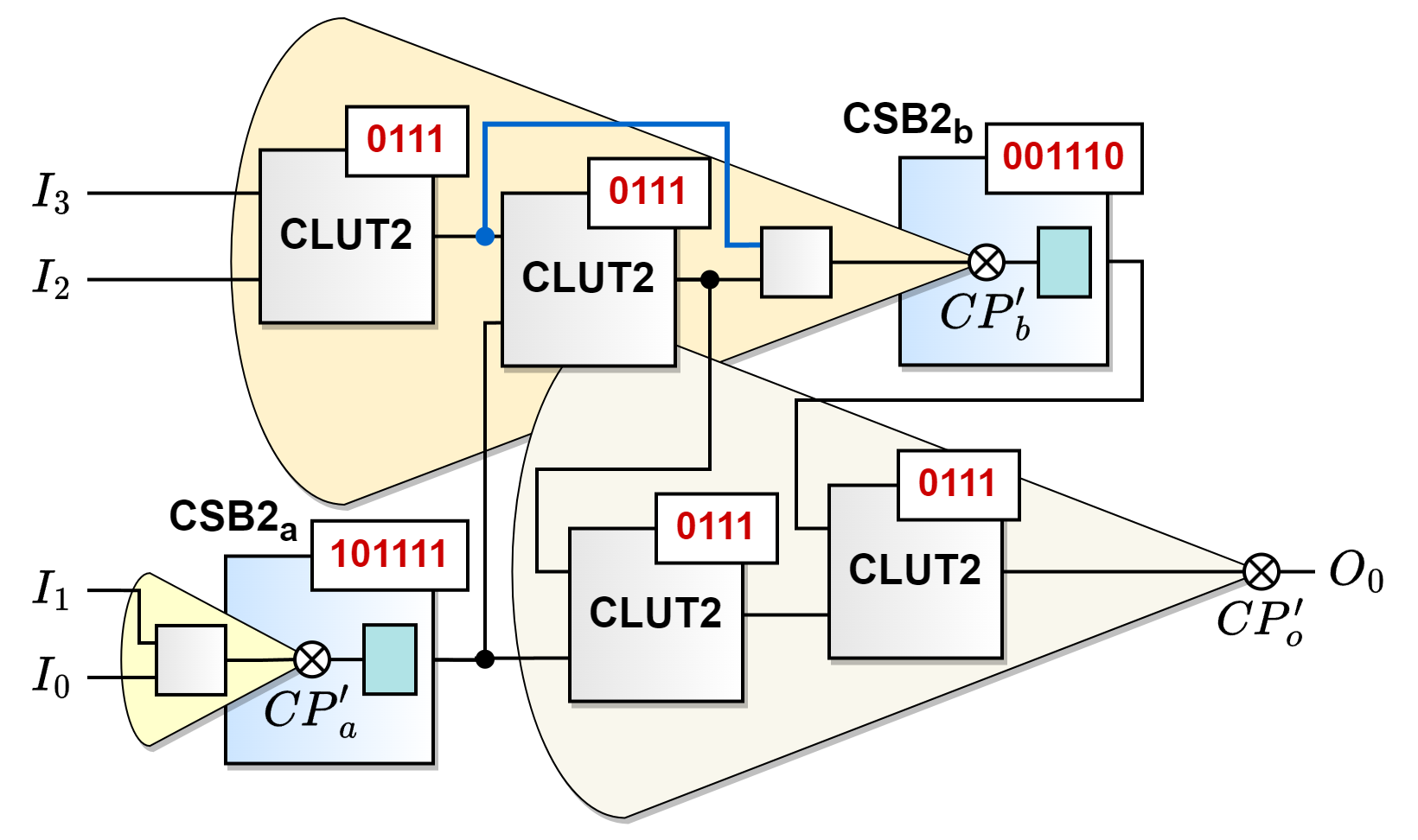}}
    \caption{Randomized CSB placement in \ciphrs ($\mathcal{RT}_{4}$): (a) Boolean logic redaction using only CLUT2s with potential locations for dummy CSBs, and (b) Updated configurable fabric with two dummy CSB2s. \textbf{CSB2\textsubscript{a}} contains a dummy FF and \textbf{CSB2\textsubscript{b}} contains a dummy CLUT and interconnect (depicted in {\color{RoyalBlue}\textbf{blue}}). \textbf{CSB2\textsubscript{a}} creates a new cut-point at ${CP}_{a}'$ and alters the fan-in cone at ${CP}_{o}'$ with the dummy FF, whereas \textbf{CSB2\textsubscript{b}} modifies the fan-in cone at ${CP}_{b}'$.}%
    \label{fig:RT4}%
\end{figure}

\ciphrs implements randomized placement of CSBs ($\mathcal{RT}_{4}$) containing dummy Boolean and sequential logic to increase the number of structural cut-points (primary outputs and FF inputs) and transform the fan-in logic cones throughout the design. Two types of dummy CSB logic are introduced in $\mathcal{RT}_{4}$ as depicted in Fig. \ref{fig:RT4}: (a) CSB with dummy sequential logic (\textbf{CSB2\textsubscript{a}}) and (b) CSB with dummy Boolean logic (\textbf{CSB2\textsubscript{b}}). For both types of dummy CSBs, dummy interconnects may be added as required, following the design topology to prevent combinational loops. The randomized placement of dummy CSBs in $\mathcal{RT}_{4}$ significantly restructures the existing logic cones in the redacted design compared to the original, while introducing new cut-points (and corresponding logic cones) with every added dummy CSB, without altering the true functionality. 

Let $n_o$ be the number of cut-points in the original design. The total number of cut-points in the redacted design can be calculated as $n_r = n_o + n_a + n_b$, where $n_a$ and $n_b$ correspond to the new cut-points added by CSBs with dummy sequential and Boolean logic, respectively. Let $\rho(n_r)$ denote the overall difficulty of predicting the true logic cone size for all $n_r$ cut-points, then the RE complexity of isolating the cut-points added in $\mathcal{RT}_{4}$ from the original cut-points by an adversary is $\binom{n_r}{n_o}.\rho(n_r)$, which increases super-exponentially even for a small number of dummy CSBs ($n_a + n_b$). 

\begin{figure}[!htbp]
    \centering
    \subfloat[]{\includegraphics[width=0.5\columnwidth]{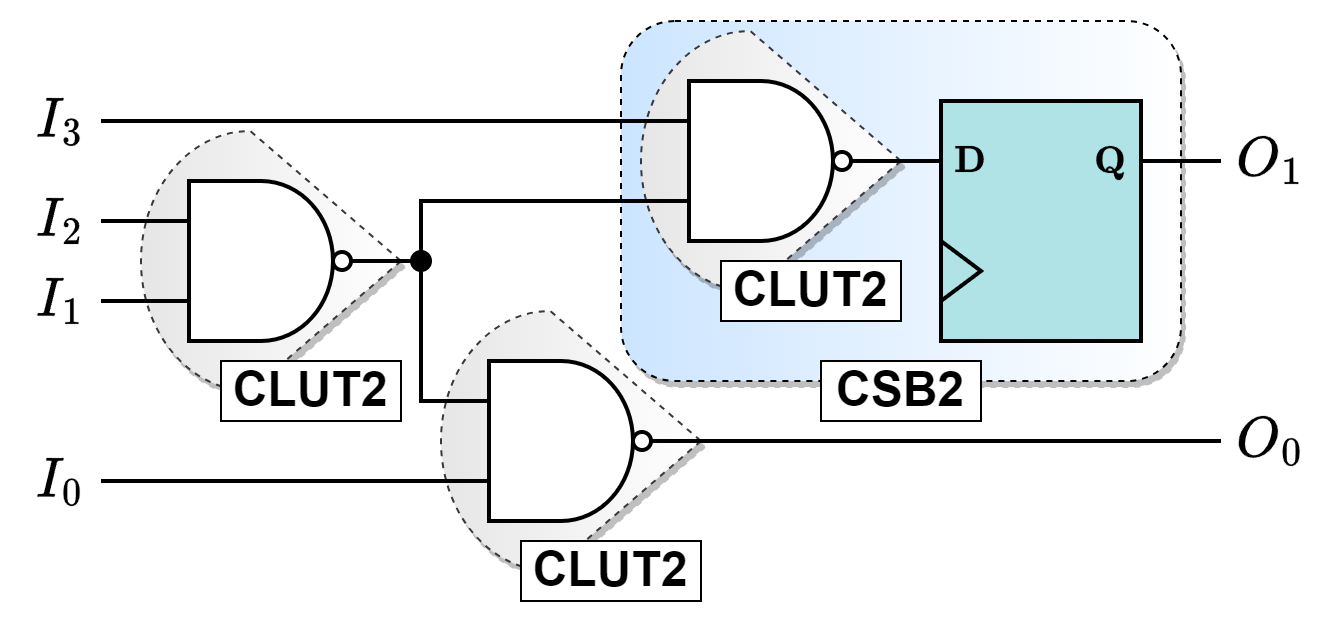}}
    \subfloat[]{\includegraphics[width=0.5\columnwidth]{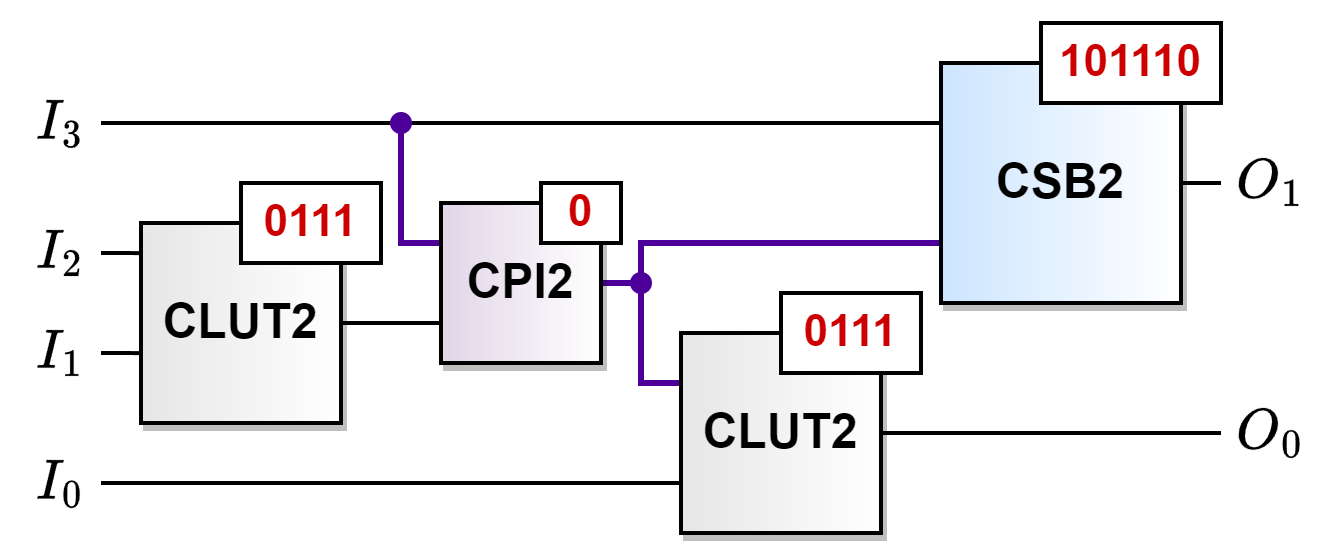}}
    \caption{Randomized CPI placement in \ciphrs ($\mathcal{RT}_{5}$): (a) Boolean logic is redacted using CLUT2 and sequential logic is redacted using CSB2, (b) CPI2 are inserted at the outputs of CLUTs/CSBs for interconnect randomization (depicted in {\color{Plum}\textbf{violet}}). The CPI placement is carefully carried out by following the design topology to avoid the creation of new combinational timing loops.}%
    \label{fig:RT5}%
\end{figure}

\subsubsection{Randomized CPI Placement}

\ciphrs utilizes randomized CPI placements ($\mathcal{RT}_{5}$) at the outputs of randomly selected CLUTs/CSBs to maximize interconnect randomization in the configurable fabric for the redacted logic. The remaining inputs of the CPIs are connected to dummy interconnects without violating the design topology to prevent the formation of combinational timing loops. The interconnect randomizations in $\mathcal{RT}_{5}$ improve the resistance against RE attacks that attempt to isolate or bypass redaction artifacts via structural analysis. Fig. \ref{fig:RT5} depicts how CPI2s are placed in the configurable fabric for interconnect randomization in \ciphr. 

For a given CLUT/CSB, the probability that its output is connected through CPI is given by $\frac{p}{r}$, where $p$ corresponds to the number of CLUT/CSB outputs with CPI(s) and $r$ is the total number of CPIs available for placement, which depends on the design and the amount security-critical logic to be redacted. The total number of possible randomized configurations is given by $\binom{p}{r}$, representing the combinatorial selection of $r$ CPIs among $p$ CLUT/CSB outputs. Higher $r$ increases interconnect randomization but may impact critical path delay, while lower $r$ satisfies delay constraints but reduces security. As the values of $p$ and $r$ increase, the complexity of the RE attack grows considerably. This, in turn, leads to a larger area overhead, which proportionally increases the overall power consumption. In \ciphr, the designer is allowed to configure the degree of CPI insertion via the redaction parameter ($\gamma$), based on their overhead budget. By default, the CPI insertion is maximized ($p = r$) to maximize interconnect randomization \cite{hipr_2025_tches}.
    
\subsection{Extension to other IP protection techniques}

The security-aware randomizations $\mathbb{RT}$ in \ciphrs can be extended to any IP protection countermeasure found to be vulnerable to RE attacks based on structural analysis of the protected design. In particular, \ciphrs can be integrated with both LUT-based \cite{bhunia2024removal,EvoLUTe_2023,hipr_2025_tches} and eFPGA-based \cite{eFPGA-CMU-2_2021, eFPGA-NYU_2021, ALICE_2022, SheLL_2023} redaction techniques, which are susceptible to RE attacks by a privileged adversary \cite{dasgupta2025latte, library_attack_2025_arXiv, knowledge_guided_attack} due to deterministic transformations and structural artifacts associated with the redacted logic. Moreover, the randomizations in $\mathbb{RT}$ can be easily extended to structured ASIC solutions like the layered IP protection methodology proposed in \cite{easic_paper} or hASICs \cite{hASIC_2023}, despite offering limited via-programmability. This flexibility makes \ciphrs a robust and adaptable security enhancement that can be incorporated into a wide range of hardware IP protection techniques to improve their resistance to RE attacks by privileged adversaries in the IC supply chain that leverage structural analysis. 

\subsection{Commercial EDA tool flow with \ciphr}

\begin{figure}[!htbp]
    \centering
    \includegraphics[width=1.0\columnwidth]{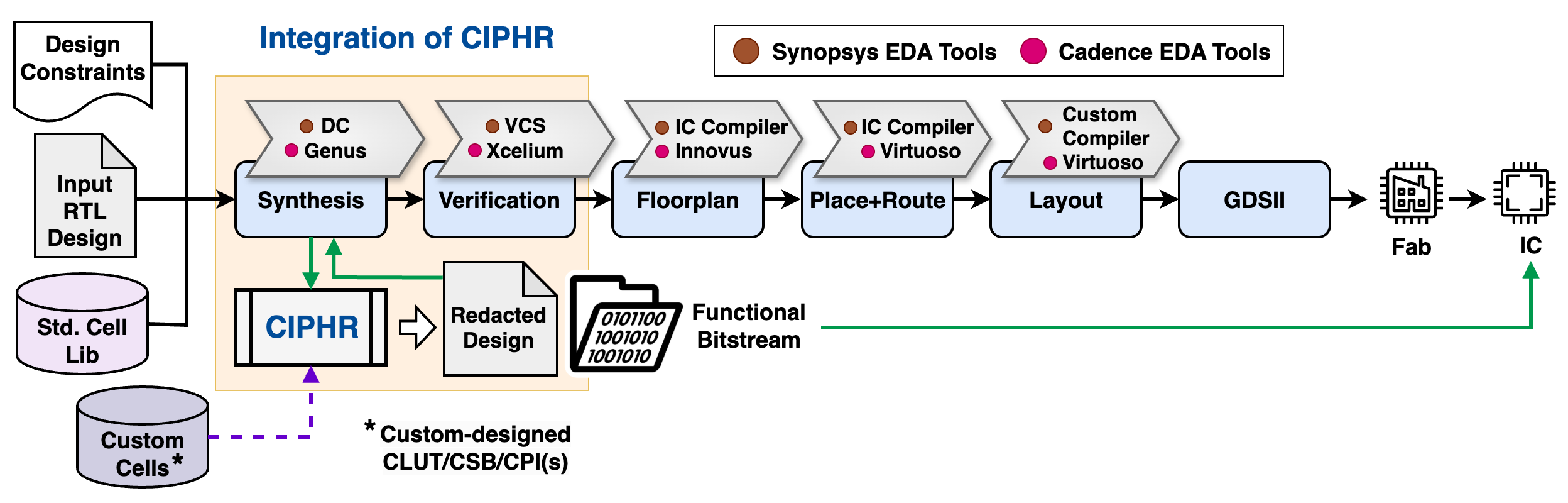}
    \caption{Integrating \ciphrs into commercial EDA tool flow. Custom-designed CLUT/CSB/CPI(s) can be used to implement the redacted logic in \ciphrs to minimize PPA overheads.}
    \label{fig:ripper_eda_flow}%
\end{figure}

\ciphrs can be seamlessly integrated with standard commercial EDA tool flow (e.g., Synopsys/Cadence), enabling fine-grained redaction of security-critical components during the synthesis phase of the IC design process. The redacted design generated by \ciphrs progresses through the remaining phases, including fabrication, while the functional bitstream is kept secure and inaccessible to untrusted entities in the supply chain. Only trusted parties with access to the bitstream can program the design, protecting the IP from confidentiality and integrity attacks. This integration streamlines the workflow, improves design flexibility, and increases resilience to RE attacks. Fig. \ref{fig:ripper_eda_flow} illustrates how \ciphrs is integrated within the synthesis and verification stages of the commercial EDA tool flow. The functional bitstream generated by \ciphrs is deployed with the IC to the end user, similar to an FPGA bitstream. The end user can restore the functionality of the redacted design with the bitstream at runtime to enable the intended operation, similar to FPGA power-up initialization.

Custom-designed CLUT/CSB/CPI(s) with transistor-level optimizations can be used by \ciphrs during the synthesis stage to realize the security-critical logic in the redacted netlist generated by \ciphrs and minimize the PPA overheads, as shown in Fig. \ref{fig:ripper_eda_flow}. The subsequent stages in the EDA tool flow are then modified to accommodate the configurable fabric using custom cells. In particular, the simulation model for the custom cells is required for the verification stage, and the physical design model (if available) can be leveraged during the place+route stage to strategically align the custom cells and optimize their placement. However, these modifications also introduce additional complexity and deviate from the standard EDA tool flow, requiring custom setups at different stages. The reliance on custom-cell libraries also increases the difficulty of bitstream verification and reduces design portability. 

\section{Transformed Design Indistinguishability}
\label{sec:tdi}

We describe how the cryptographic property of indistinguishability can be extended to quantify the design transformations introduced by security countermeasures such as hardware redaction. The transformed design indistinguishability ($\mathbf{TDI}$) can be categorized as follows:

\subsection{Functional Indistinguishability}

A privileged adversary from our threat model can try to extract the transformed design features and employ statistical analysis to isolate the original functionality from these features. To thwart such an adversary and ensure functional indistinguishability ($\mathbf{TDI}_{F}$) in the redacted designs, it is necessary to increase the possible functional space to such an extent that statistical analysis for RE becomes infeasible. 

For the fine-grain redaction in \ciphr, the total number of functions possible for a given $n$-input CLUT is $2^{2^{n}}$. Of these, the number of functions that depend on all $n$ inputs can be calculated using the following recurrence relation:
\begin{dmath} \label{eq:tdi_f}
    {\mathcal{F}_{n}} = 2^{2^{n}} - ({\mathcal{F}_{n-1}} + 2); \hspace{1em} {\mathcal{F}_{0}\leftarrow0}, {\mathcal{F}_{1}\leftarrow2}.
\end{dmath}

As $\mathcal{F}_{n}$ is proportional to the CLUT size $n$, larger CLUTs can help increase $\mathbf{TDI}_{F}$. In \ciphr, the randomized input space expansion ($\mathcal{RT}_{2}$) allows the CLUT input sizes in the configurable fabric to be increased, while the randomized input reordering and output inversion ($\mathcal{RT}_{3}$) combined with the dummy Boolean logic from randomized CSB insertion ($\mathcal{RT}_{4}$) expand the CLUT functionality distribution, for the same redacted logic.

\subsection{Structural Indistinguishability}

In our threat model, a privileged adversary can perform RE using structural analysis of the transformed design to identify any artifacts corresponding to the security countermeasures and remove them from the design. To prevent this, each redacted design must possess a significant degree of structural indistinguishability ($\mathbf{TDI}_{S}$). 

For a gate-level netlist or hypergraph, the primary outputs ($PO$) and data inputs of FFs ($pseudo$-$PO$) constitute the set of structural cut-points ($\mathcal{CP}$) in the design. For a given cut-point ${CP}_{i} \in \mathcal{CP}$, we can define the $\mathbf{TDI}_{S}$ metric from its structural features as follows: 

\begin{dmath} \label{eq:tdi_s}
    {\mathbf{TDI}_{S}} = w_1\cdot\mathbf{FI}_{size} + w_2\cdot\mathbf{FO}_{size} + w_3\cdot\mathbf{FI}_{gates} + w_4\cdot\mathbf{FI}_{drivers}
\end{dmath}

where, $\mathbf{FI}_{size}$ is the size of fan-in-cone at ${CP}_{i}$, $\mathbf{FO}_{size}$ is the size of fan-out-cone from ${CP}_{i}$, $\mathbf{FI}_{gates}$ is the number of gates in the fan-in-cone and $\mathbf{FI}_{drivers}$ the number of inputs driving the fan-in-cone, which include primary inputs ($PI$) and FF outputs ($pseudo$-$PI$). The $\mathbf{TDI}_{S}$ metric can be configured using the scaling weights $\{w_1,w_2,w_3,w_4\}$. In \ciphr, the randomized CLUT mapping ($\mathcal{RT}_{1}$) combined with structural transformations due to the randomized CSB and CPI placements ($\mathcal{RT}_{4}$ and $\mathcal{RT}_{5}$) helps increase $\mathbf{TDI}_{S}$ for the redacted logic.
    
\section{Results and Analysis}
\label{sec:result}

\subsection{Experimental Setup}

We use multiple open-source benchmarks from the \textit{ISCAS89} \cite{iscas89}, \textit{ITC99} \cite{itc99}, and \textit{MIT-CEP} \cite{MITCEP} suites to evaluate the functional and structural indistinguishability in redacted designs generated by \ciphr. Synopsys Design Compiler (V-2023.12-SP5) is used in the synthesis stages to generate the gate-level netlists and the associated PPA metrics for the NanGate 15nm open cell library \cite{si2openpdk2025}. The functional bitstream for the configurable fabric is obtained via simulation using Synopsys VCS (V-2023.12). Logic Equivalence Checking (LEC) using Cadence Conformal (21.10-s300) is used to generate similarity metrics. Our experiments were performed on a Red Hat Enterprise Linux Server 7.9 server with AMD® Epyc 7713 64-core processor and 1007.6 GiB of memory.

\subsection{Evaluation Benchmarks}

Table \ref{tab:overheads} lists the open-source benchmarks used for the security evaluation of the redacted designs in \ciphrs and their specifications for NanGate 15nm. The evaluation benchmarks (\textbf{OD$i$}, $i \in \{1,2,3,4,5,6\}$) are selected from different benchmark suites, with an average cell count of $\sim$7.7 K, to ensure that the proposed randomizations in $\mathbb{RT}$ are robust and scalable. It should be noted that although \ciphrs allows fine-grain redaction at the gate level, a single module instance for each evaluation benchmark is 100\% redacted to enable the comparison with existing fine-grain and coarse-grain redaction techniques. For every \textbf{OD$i$}, we generate 4 redacted variants \textbf{RD$i$-F$j$} ($j \in \{0,1,2,3\}$) and 5 redacted variants \textbf{RD$i$-S$k$} ($k \in \{0,1,2,3,4\}$) to analyze both functional ($\mathbf{TDI}_{F}$) and structural indistinguishability ($\mathbf{TDI}_{S}$), respectively, by varying the redaction parameters in $\gamma$. The non-randomized baseline variants (\textbf{RD$i$-F$0$}~$\equiv$~\textbf{RD$i$-S$0$}) variants lack security-aware randomizations ($\mathbb{RT}$) and correspond to existing fine-grain redaction techniques~\cite{hipr_2025_tches}.

\begin{table*}[!htbp]
    \centering
    \caption{Overhead results for redacted designs (\textbf{RD$i$}, $i \in \{1,2,3,4,5,6\}$) generated by \ciphr.}
    \label{tab:overheads}
    \resizebox{1.0\textwidth}{!}{
    \begin{tabular}{c|c|c|cccc|ccc|ccc|ccc}
    \hline
    \multirow{3}*{\textbf{Test \#}} & \textbf{Redacted} & \textbf{\% Logic} & 
    \multicolumn{4}{c|}{\multirow{2}*{\textbf{IP-level metrics for} \textbf{OD$i$}}\ddg} & 
    \multicolumn{3}{c|}{\textbf{Overheads* w/o $\mathbb{RT}$}} &
    \multicolumn{6}{c}{\textbf{Average Overheads* with $\mathbb{RT}$ in \ciphr}\cdt} \\ \cline{11-13} \cline{14-16}
    {} & \textbf{Module} & \textbf{Redacted} & 
    {} & {} & {} & {} & 
    \multicolumn{3}{c|}{(\textbf{RD$i$-F$0$} $\equiv$ \textbf{RD$i$-S$0$})\bstars} &
    \multicolumn{3}{c|}{$\langle$\textbf{\textbf{RD$i$-F$j$}}$\rangle$} &
    \multicolumn{3}{c}{$\langle$\textbf{\textbf{RD$i$-S$k$}}$\rangle$} \\ \cline{4-7} \cline{8-10} \cline{11-13} \cline{14-16}
    {} & \textbf{(\textbf{OD$i$})} & \textbf{(by Area)} &
    $\mathbf{\left|\mathbf{Cells}\right|}$ & \textbf{Area} & \textbf{Delay} & \textbf{Power} &
    \textbf{Area} & \textbf{Delay} & \textbf{Power} &
    \textbf{Area} & \textbf{Delay} & \textbf{Power} &
    \textbf{Area} & \textbf{Delay} & \textbf{Power} \\
    \hline 
    \textbf{RD${1}$} & $\textbf{B21}$    & 100\% & 6700	& 3305.67	& 1143.3	& 0.91 & 2.61x   & 0.01x	& 8.35x		& 6.93x	 & 0.00x	& 24.73x & 4.29x	& 0.00x	& 15.37x	 \\
    \textbf{RD${2}$} & $\textbf{B22}$    & 100\% & 9949	& 4919.67	& 1143.3	& 1.41 & 2.63x   & 0.00x	& 8.07x		& 6.96x	 & 0.00x	& 24.01x & 4.30x	& 0.00x	& 15.01x	\\ 
    \hdashline
    \textbf{RD${3}$} & $\textbf{S38417}$ & 100\% & 7376	& 3515.74	& 1308.11	& 1.85 & 3.46x   & 0.79x	& 5.42x	    & 8.29x	 & 0.63x	& 13.16x & 5.37x	& 0.60x	& 8.58x \\
    \textbf{RD${4}$} & $\textbf{S38584}$ & 100\% & 9520	& 3576.84	& 1216.53	& 1.74 & 3.49x   & 0.56x	& 5.60x		& 8.84x	 & 0.31x	& 14.47x & 5.61x	& 0.47x	& 9.14x \\ 
    \hdashline
    \textbf{RD${5}$} & $\textbf{MD5}$    & 100\% & 6541	& 2696.48	& 874.94	& 0.69 & 4.01x	 & 0.00x    & 14.99x	& 13.00x & 0.00x	& 46.01x & 8.39x	& 0.00x	& 28.62x	 \\
    \textbf{RD${6}$} & $\textbf{SHA256}$ & 100\% & 5995	& 3194.39	& 1264.61	& 1.78 & 4.44x   & 0.44x	& 6.81x		& 11.27x & 0.01x	& 17.19x & 7.10x	& 0.17x	& 11.07x	 \\
    \hline
    \textbf{Average} &  & \textbf{100\%}
    & \textbf{7681}		& \textbf{3534.80} & \textbf{1158.47}		& \textbf{1.40}
    	& \textbf{3.44x} & \textbf{0.30x}		& \textbf{8.21x}
    	& \textbf{9.22x} & \textbf{0.16x}		& \textbf{23.26x}
    	& \textbf{5.85x} & \textbf{0.21x}	& \textbf{14.63x} \\
    \hline 
    \end{tabular}%
    }
    \scriptsize
    \raggedright\\
    \ddg \textbf{Units: Power ($mW$), Delay ($ps$), Area ($\mu$$m^{2}$).} \bstars \textbf{Non-randomized variant, represents prior art~\cite{hipr_2025_tches}.}
    \cdt \textbf{$\mathbf{j \in \{1,2,3\}}$, $\mathbf{k \in \{1,2,3,4\}}$.} \\
    \textbf{* Overheads are normalized and reported as x times original.}
\end{table*}

\subsection{Overhead Analysis}

For calculating PPA overheads, all synthesis stages are constrained to an operating clock frequency of 500 MHz, and the gate-level netlists are mapped to the full NanGate 15nm (typical) standard cell library. Table \ref{tab:overheads} reports the overheads obtained for the redacted variants for $\mathbf{TDI}_{F}$ and $\mathbf{TDI}_{S}$ in \ciphr. The reported overheads are calculated using the absolute values of the PPA metrics from synthesis logs (not the NAND2 gate-equivalent) and are normalized using the original PPA values. The overhead values reported for the redacted variants with security-aware randomizations ($\mathbb{RT}$) are separately averaged over all the \textbf{RD$i$-F$j$} and \textbf{RD$i$-S$k$} variants. 

We can observe from Table \ref{tab:overheads} that on average, the PPA overheads for the non-randomized baseline variants \textbf{RD$i$-F$0$} and \textbf{RD$i$-S$0$} are noticeably lower than the redacted variants with $\mathbb{RT}$. For example, the area overheads are almost tripled from 3.44x (baseline) to 9.22x in the case of $\mathbf{TDI}_{F}$ and doubled to 5.85x in the case of $\mathbf{TDI}_{S}$, with a proportional increase in power overheads. The higher overheads for \textbf{RD$i$-F$j$} compared to \textbf{RD$i$-S$k$} are due to the skewing of redaction parameters to include more CLUTs with larger input sizes to increase $\mathbf{TDI}_{F}$, as discussed later in Section \ref{subsubsec:tdi_f}, and the average overheads obtained for \textbf{RD$i$-S$k$} better represent the overall redaction implemented in \ciphr. The delay overheads for all variants are less than 1x, implying that redacted designs do not violate timing constraints and meet the slack.

\subsubsection{Comparison with state-of-the-art}

We compare the overheads obtained from the proposed fine-grain redaction in \ciphrs with fine-grain \cite{Custom-LUT_2019, EvoLUTe_2023, hipr_2025_tches} and coarse-grain \cite{eFPGA-CMU-2_2021, ALICE_2022, SheLL_2023} redaction techniques. Table \ref{tab:compare} compares the average area overheads and bitstream sizes for various redaction methodologies. For simplifying the analysis, we chose state-of-the-art techniques \textit{EvoLUTe} \cite{EvoLUTe_2023} and \textit{SheLL} \cite{SheLL_2023} as representatives of the best fine-grain and coarse-grain redaction approaches, respectively, and compare them to the non-randomized \ciphrs redaction as the baseline. Since evaluation benchmarks and redacted IP sizes differ significantly, we scale the reported overheads to account for the redaction of an equivalent amount of logic as \ciphr, measured using NAND2 gate-equivalent cells. The area overheads reported for \textit{EvoLUTe} are lower than actual values since the bitstream storage area is ignored and the target cell library is unspecified. For a fair comparison, the overhead for the eFPGA equivalent (as reported) is used to determine the redacted IP size and estimate the actual overhead and bitstream size.

Table \ref{tab:compare} shows that the proposed security-aware fine-grain redaction in \ciphrs can improve the overhead by more than an order of magnitude for the same amount of logic redacted compared to the state-of-the-art. The area overheads (scaled) are reduced from 97.25x for coarse-grain (eFPGA-based) and 128.67x for fine-grain (LUT-based) techniques to 9.22x in \ciphrs \textbf{(RD$i$-F$j$)} and to 5.85x in \ciphrs \textbf{(RD$i$-S$k$)} and 3.54x without the randomizations with $\mathbb{RT}$. The bitstream sizes for \ciphrs are also lower than the state-of-the-art by an order of magnitude, which implies faster configuration times along with a lower overhead from the bitstream storage and encryption/decryption logic.

\begin{table}[!htbp]
    \centering
    \caption{Area overhead and bitstream size comparison between existing redaction techniques and \ciphr.}
    \label{tab:compare}
    \resizebox{\columnwidth}{!}{%
    \begin{tabular}{c|c|cc|cc}
    \hline
    \multirow{2}*{\textbf{Technique}} &
    \textbf{Redacted} &
    \multicolumn{2}{c|}{\textbf{Area Overheads\textsuperscript{b}}} &
    \multicolumn{2}{c}{\textbf{Bitstream Size\textsuperscript{c}}} \\ \cline{3-4} \cline{5-6}
    {} &
    \textbf{IP Size\textsuperscript{a}} &
    \textbf{Reported} & 
    \textbf{Scaled*} &
    \textbf{Reported} &
    \textbf{Scaled*} \\
    \hline
    \textbf{SheLL} \cite{SheLL_2023} & 273 & 1.48x & 97.25x & 16 K & 995 K  \\
    \textbf{EvoLUTe} \cite{EvoLUTe_2023} & 460 & 3.29x & 128.67x & 14 K & 524 K \\
    \ciphr\bstars\cdt~\textbf{(\underline{No} $\mathbb{RT}$)} & 17979 & 3.44x & 3.44x & 32 K & 32 K  \\
    \ciphr\bstars\ddg~\textbf{(RD$i$-F$j$)} & 17979 & 9.22x & 9.22x & 82 K & 82 K \\
    \ciphr\bstars\ddg~\textbf{(RD$i$-S$k$)} & 17979 & 5.85x & 5.85x & 52 K & 52 K \\
    \hline 
    \end{tabular}%
    }
    \scriptsize
    \raggedright\\
    \bstars \textbf{Current work.} \ddg \textbf{Including randomizations $\mathbb{RT}$ for $\mathbf{TDI}_{F}$ and $\mathbf{TDI}_{S}$.} \\
    \textbf{* Scaled to match the amount of logic redacted by \ciphr.} \\
    \textbf{\textsuperscript{a} NAND2 gate equivalent (GE).} \textbf{\textsuperscript{b} Normalized to original area (1x).} \\
    \textbf{\textsuperscript{c} Multiples of 1 K (1000) bits.} \cdt \textbf{Represents prior art~\cite{hipr_2025_tches}.}
\end{table}

\subsection{Security Evaluation}

\subsubsection{Functional Indistinguishability}
\label{subsubsec:tdi_f}

In order to maximize $\mathbf{TDI}_{F}$ in \ciphr, the objective during redaction is to increase both the count and the frequency of unique functions (or bitstreams) realized by the configurable fabric. Since the number of unique functions possible in CLUT4 $>$ CLUT3 $>$ CLUT2, we attempt to increase $\mathbf{TDI}_{F}$ by implementing the same redacted logic using larger CLUTs with a wider range of functionalities. For each redacted IP \textbf{RD$i$} ($i \in \{1,2,3,4,5,6\}$), we generate 4 distinct variants \textbf{RD$i$-F$j$} ($j \in \{0,1,2,3\}$) under different tool settings. The \textbf{RD$i$-F$0$} variant does not contain the randomizations in $\mathbb{RT}$ and serves as a baseline for comparison. The remaining variants \textbf{RD$i$-F$\{1,2,3\}$} are generated by increasing the minimum allowed CLUT size ($\gamma_{min}$) from 2 to 4, and randomized using the seed ($\theta$).

\begin{table*}[!ht]
    \centering
    \caption{CLUT counts and bitstream size variation under different redaction parameter ($\gamma$) and random seed ($\theta$) settings in \ciphr.}
    \label{tab:clut_cov}
    \resizebox{\textwidth}{!}{
    \begin{tabular}{c|cc|cc|cc|cc}
    \hline
    \multirow{2}*{\textbf{Test \#}} & 
    \multicolumn{2}{c|}{\textbf{RD$i$-F$0$} ($\gamma_{0}:\gamma_{min}\leftarrow2,\theta_{0}$)} &
    \multicolumn{2}{c|}{\textbf{RD$i$-F$1$} ($\gamma_{1}:\gamma_{min}\leftarrow2,\theta_{1},\mathbb{RT}$)} &
    \multicolumn{2}{c|}{\textbf{RD$i$-F$2$} ($\gamma_{2}:\gamma_{min}\leftarrow3,\theta_{2},\mathbb{RT}$)} &
    \multicolumn{2}{c}{\textbf{RD$i$-F$3$} ($\gamma_{3}:\gamma_{min}\leftarrow4,\theta_{3},\mathbb{RT}$)} \\ \cline{2-3} \cline{4-5} \cline{6-7} \cline{8-9}
    {} &
    \cluts\ddg & \bits &
    \cluts\ddg & \bits &
    \cluts\ddg & \bits &
    \cluts\ddg & \bits \\
    \hline 
    \textbf{RD$1$-F$j$} & {3203, 1289, 38}	& 23732-bit	& {722, 66, 3742}	& 63288-bit	& {0, 2547, 1983}	& 52104-bit	& {0, 0, 4530}	& 72480-bit  \\
    \textbf{RD$2$-F$j$} & {4707, 2045, 37}	& 35780-bit	& {1082, 191, 5516} & 94112-bit	& {0, 3816, 2973}	& 78096-bit	& {0, 0, 6789}	& 108624-bit  \\
    \textbf{RD$3$-F$j$} & {3785, 1731, 65}	& 30028-bit	& {882, 255, 4444}	& 76672-bit	& {0, 3158, 2423}	& 64032-bit	& {0, 0, 5581}	& 89296-bit  \\
    \textbf{RD$4$-F$j$} & {4351, 1723, 51}	& 32004-bit	& {971, 78, 5076}	& 85724-bit	& {0, 3458, 2667}	& 70336-bit	& {0, 0, 6125}	& 98000-bit  \\
    \textbf{RD$5$-F$j$} & {4201, 1598, 30}	& 30068-bit	& {931, 3271, 1627} & 55924-bit	& {0, 3271, 2558}	& 67096-bit	& {0, 0, 5829}	& 93264-bit  \\
    \textbf{RD$6$-F$j$} & {4877, 2177, 33}	& 37452-bit	& {1126, 241, 5720} & 97952-bit	& {0, 3992, 3095}	& 81456-bit	& {0, 0, 7087}	& 113392-bit  \\
    \hline
    \textbf{Average} & \textbf{{4188, 1761, 43}} & \textbf{31511-bit} & \textbf{{953, 684, 4355}} & \textbf{78946-bit} &	\textbf{{0, 3374, 2617}} & \textbf{68854-bit} & \textbf{{0, 0, 5991}} & \textbf{95843-bit} \\
    \hline 
    \end{tabular}
    }
    \scriptsize
    \raggedright\\ 
    \cdt \textbf{$\mathbf{j \in \{0,1,2,3\}}$.} \ddg \textbf{$\mathbf{CL2: CLUT2, CL3: CLUT3, CL4: CLUT4}$.}
\end{table*}

\begin{figure*}[!htbp]
    \centering
    \includegraphics[width=\textwidth]{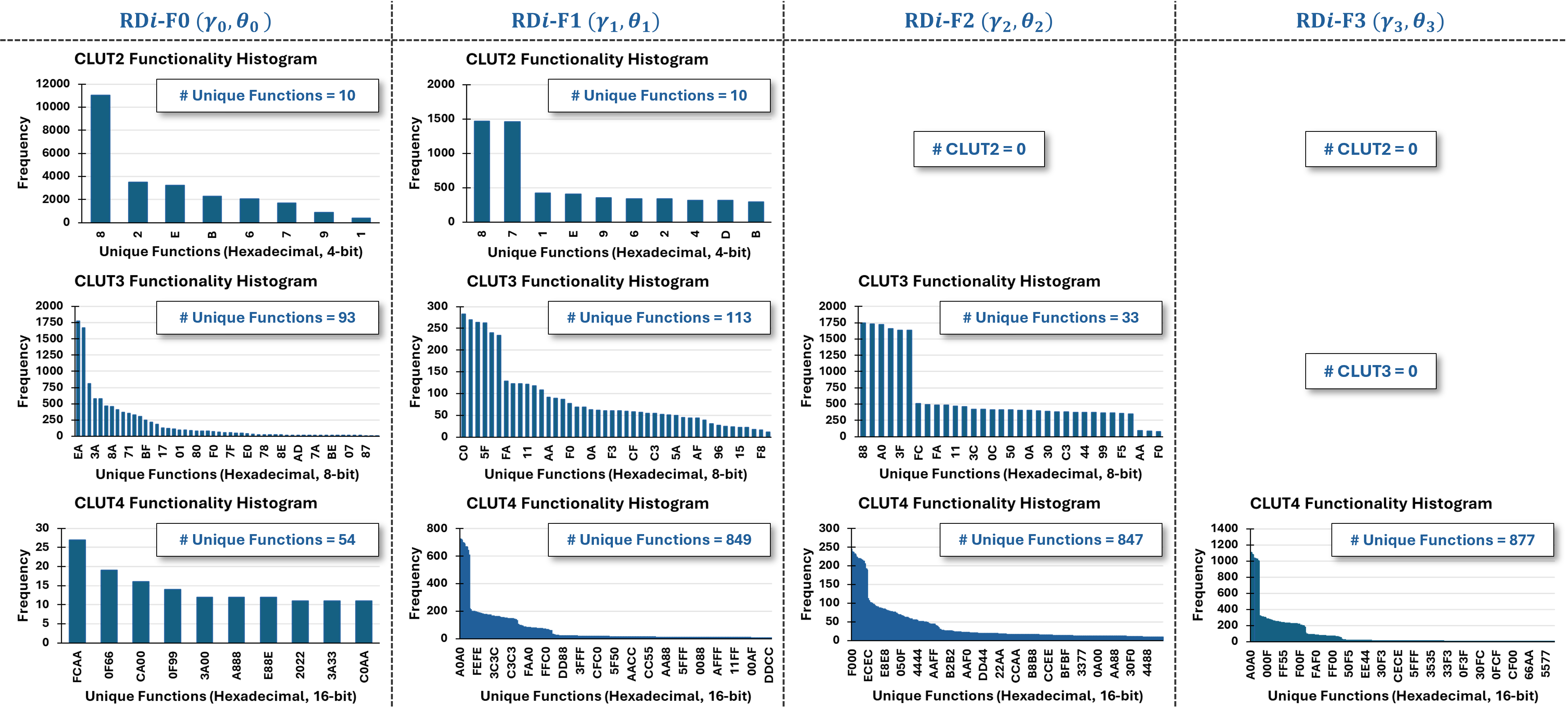}
    \caption{Cumulative frequency distribution for unique CLUT2/CLUT3/CLUT4 functions across redacted design variants generated by \ciphr.}
    \label{fig:clut_hist}
\end{figure*}

Table \ref{tab:clut_cov} shows the CLUT counts and bitstream sizes for redacted IP variants under different $\gamma$ and $\theta$ settings in \ciphr. Fig. \ref{fig:clut_hist} displays the CLUT cumulative functionality distributions of these variants corresponding to the same \ciphrs tool settings. The baseline variants \textbf{RD$i$-F$0$} (tool settings: $\gamma_{0}$, $\theta_{0}$) across all 6 variants demonstrate fewer unique functions and non-uniform frequency distributions skewed towards a small subset of functionalities. This occurs due to the smaller CLUT sizes (with fewer Boolean logic functions possible), combined with the absence of any randomizations from $\mathbb{RT}$, resulting in low $\mathbf{TDI}_{F}$. Adding randomizations from $\mathbb{RT}$ to the \textbf{RD$i$-F$1$} variants (tool settings: $\gamma_{1}$, $\theta_{1}$) with $\gamma_{min}=2$ increases the number of unique functions and makes the frequency distribution less skewed and more uniform for every CLUT size, resulting in higher $\mathbf{TDI}_{F}$. As $\gamma_{min}$ increases to 3 for the \textbf{RD$i$-F$2$} (tool settings: $\gamma_{2}$, $\theta_{2}$) and then to 4 for the \textbf{RD$i$-F$3$} (tool settings: $\gamma_{3}$, $\theta_{3}$), both these variants exhibit higher counts of larger CLUTs and demonstrate frequency distributions with greater uniformity due to a more equitable occurrence of various functions, although the number of unique functions for CLUT4 remains similar across the \textbf{RD$i$-F$\{1,2,3\}$} variants with randomizations from $\mathbb{RT}$, implying a saturation in the number of unique Boolean functions possible across the evaluation benchmarks. The combination of increased uniformity and the exponential increase in the number of possible functionalities in larger CLUTs maximizes $\mathbf{TDI}_{F}$ across the redacted IP variants \textbf{RD$i$-F$\{1,2,3\}$}, highlighting the effectiveness of randomization by \ciphrs in enhancing security.

\textbf{Balancing $\mathbf{TDI}_{F}$ vs Overheads:} Larger CLUTs (coupled with $\mathbb{RT}$) are desirable during redaction to maximize $\mathbf{TDI}_{F}$, but they also incur higher PPA overheads. For real-world applications, the degree of $\mathbf{TDI}_{F}$ in the redacted logic must be balanced against any overhead constraints specified by the user while satisfying the minimum security requirements (in terms of bitstream size). As shown in Table \ref{tab:clut_cov}, the average bitstream size increases from 32 K bits for \textbf{RD$i$-F$0$} (baseline) to 79 K bits for \textbf{RD$i$-F$1$}, then decreases to 69 K bit for \textbf{RD$i$-F$2$}, and increases to 96 K bits for \textbf{RD$i$-F$3$}. Although \textbf{RD$i$-F$3$} is more secure with higher $\mathbf{TDI}_{F}$, \textbf{RD$i$-F$2$} offers a better balance with lower overheads and high enough $\mathbf{TDI}_{F}$. The security-aware fine-grain redaction in \ciphrs enables easy configuration of redaction parameters ($\gamma$) to meet security requirements under user-specified overhead constraints.

\begin{figure*}[!htbp]
    \centering
    \includegraphics[width=\textwidth]{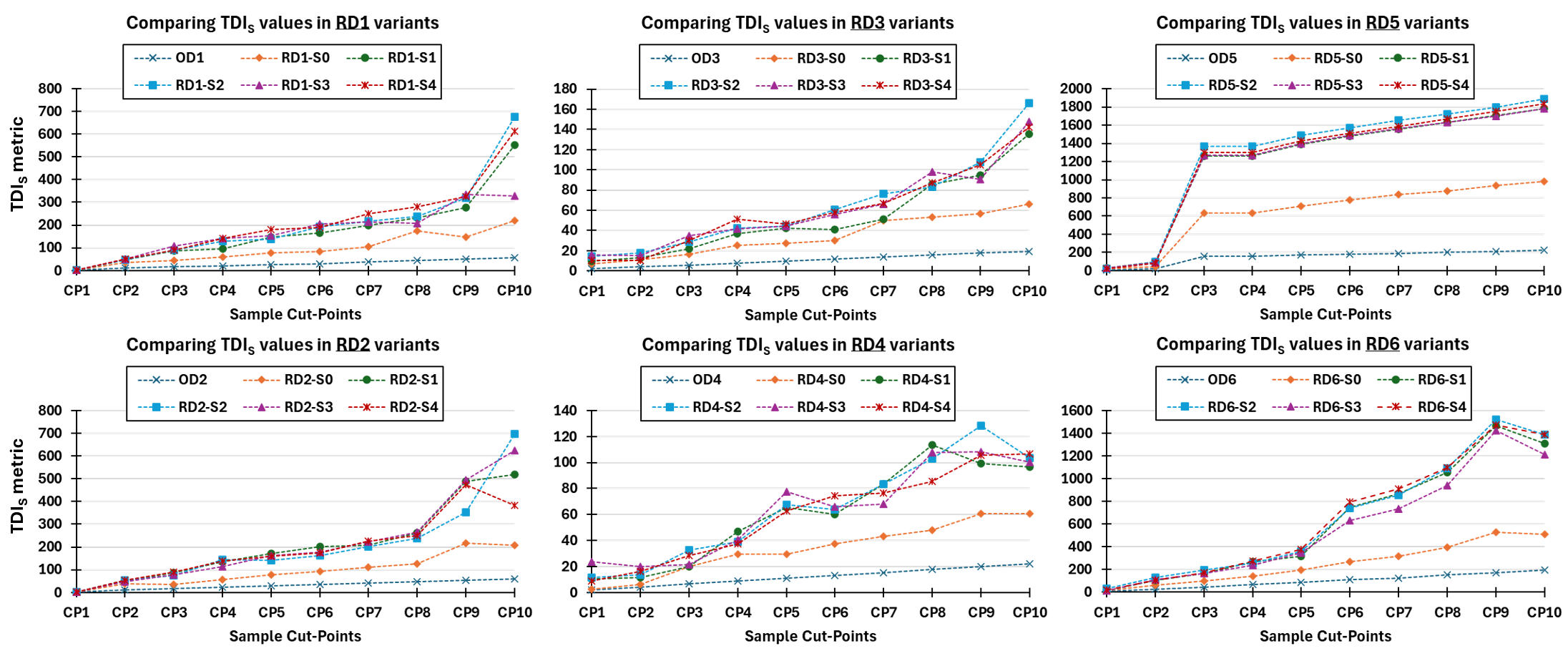}
    \caption{Plots depicting the variation in $\mathbf{TDI}_{S}$ metric values for the same sample cut-points across redacted design variants generated by \ciphr.}
    \label{fig:s_metric_variation}
\end{figure*}

\subsubsection{Structural Indistinguishability}

Each randomized transformation in $\mathbb{RT}$ aims to maximize $\mathbf{TDI}_{S}$ across the redacted IP variants. For each redacted IP \textbf{RD$i$} ($i \in \{1,2,3,4,5,6\}$), we vary the tool settings in \ciphrs to generate 5 different variants \textbf{RD$i$-S$j$} ($j \in \{0,1,2,3,4\}$). The randomizations in $\mathbb{RT}$ are not applied to the redacted variant \textbf{RD$i$-S$0$}, which serves as a baseline for comparison. The 4 remaining variants \textbf{RD$i$-S$\{1,2,3,4\}$} are substantially randomized using different values of $\theta$. After generating the redacted variants, the cut-points $\mathcal{CP}$ for each variant \textbf{RD$i$-S$j$} and the original IP \textbf{OD$i$} are identified, and $\mathbf{TDI}_{S}$ metric is then calculated for each cut-point using Eq. \ref{eq:tdi_s}.

\begin{figure*}[!htbp]
    \centering
    \includegraphics[width=\textwidth]{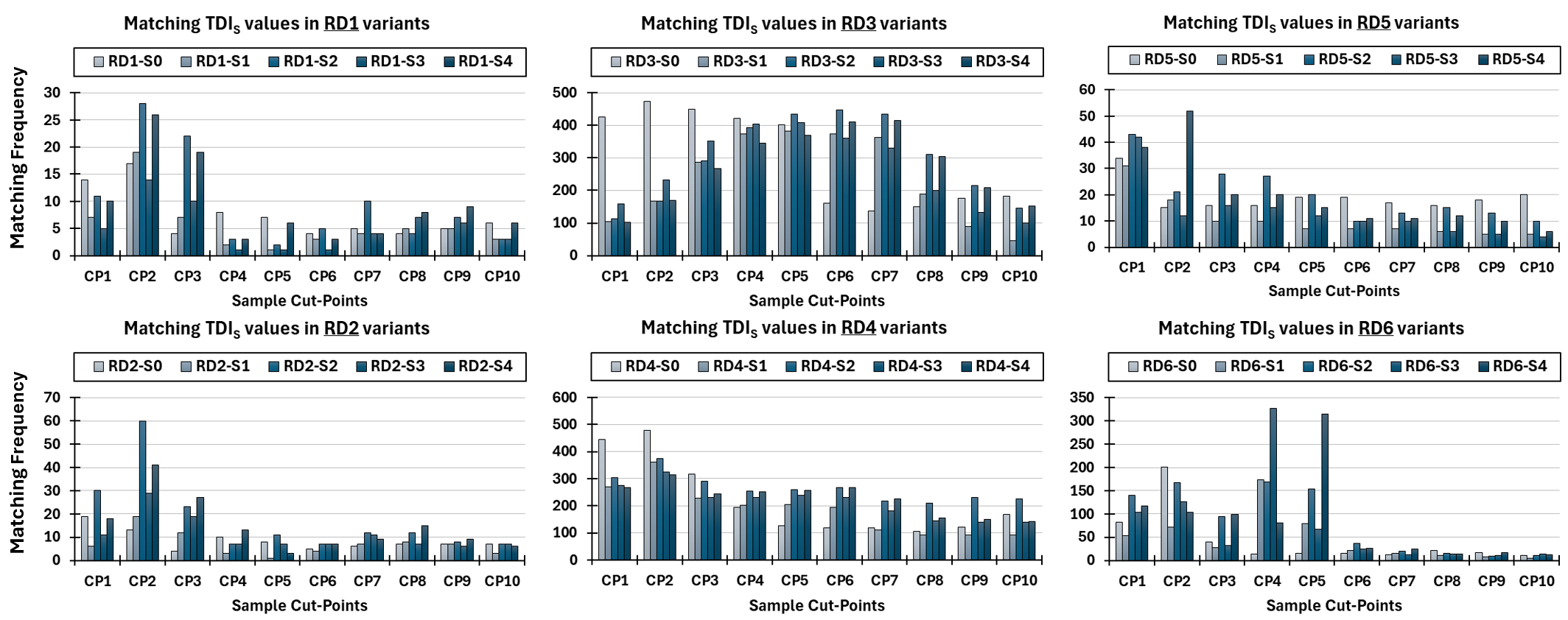}
    \caption{Plots illustrating the matching $\mathbf{TDI}_{S}$ metric values for sample cut-points from redacted design variants generated by \ciphr.}
    \label{fig:s_metric_matches}
\end{figure*}

Fig. \ref{fig:s_metric_variation} illustrates how $\mathbf{TDI}_{S}$ metric values vary across redacted IP variants for the same set of sample cut-points $\mathcal{CP}_{sample} \subseteq \mathcal{CP}$. To ensure proper coverage, the cut-points selected from $\mathcal{CP}_{sample}$ for plotting in Fig. \ref{fig:s_metric_variation} correspond to 10 equidistant samples between the minimum and maximum values of the $\mathbf{TDI}_{S}$ metric in the original IP \textbf{OD$i$}, determined and plotted separately for each redacted IP \textbf{RD$i$} ($i \in \{1,...,6\}$). It can be observed from Fig. \ref{fig:s_metric_variation} that for the same sample cut-point \textbf{CP}$x$ ($x \in \{1,...,10\}$) from $\mathcal{CP}_{sample}$ the $\mathbf{TDI}_{S}$ metric values across all redacted variants \textbf{RD$i$-S$j$} are considerably higher compared to the original IP (\textbf{OD$i$}), including the baseline variants \textbf{RD$i$-S$0$}. However, when the randomizations in $\mathbb{RT}$ are introduced, the $\mathbf{TDI}_{S}$ metric values increase significantly with noticeable deviations from the non-randomized baseline \textbf{RD$i$-S$0$}, demonstrating the degree of structural variation introduced into the redacted IP variants. For every redacted IP \textbf{RD$i$-S$j$}, we observe similar trends for randomized variations in the $\mathbf{TDI}_{S}$ metric values for its sample cut-points, although the actual magnitude of the $\mathbf{TDI}_{S}$ metric varies between the redacted IPs, implying that the security-aware randomizations in $\mathbb{RT}$ are design-independent.

Fig. \ref{fig:s_metric_matches} shows the frequencies of matching $\mathbf{TDI}_{S}$ metric values across the redacted IP variants \textbf{RD$i$-S$j$} for the same set of  sample cut-points $\mathcal{CP}_{sample}$, plotted separately for each redacted IP \textbf{RD$i$} ($i \in \{1,...,6\}$). Each cut-point \textbf{CP}$x$ ($x \in \{1,...,10\}$) with a matching value of the $\mathbf{TDI}_{S}$ metric between two variants \textbf{RD$i$-S$j$} corresponds to a false positive in terms of isolating a specific cut-point by an adversary trying to replicate the structural analysis for RE. The high values of the matching frequencies for the $\mathbf{TDI}_{S}$ metric, as observed from the separate plots for every redacted IP \textbf{RD$i$} in Fig. \ref{fig:s_metric_matches}, allow us to infer that the security-aware fine-grain redaction in \ciphrs provides an effective anti-RE countermeasure by introducing significant structural variations through the various randomizations proposed in $\mathbb{RT}$.

\begin{figure*}[!htbp]
    \centering
    \subfloat[\ciphrs heatmaps.]{\includegraphics[width=0.7\textwidth]{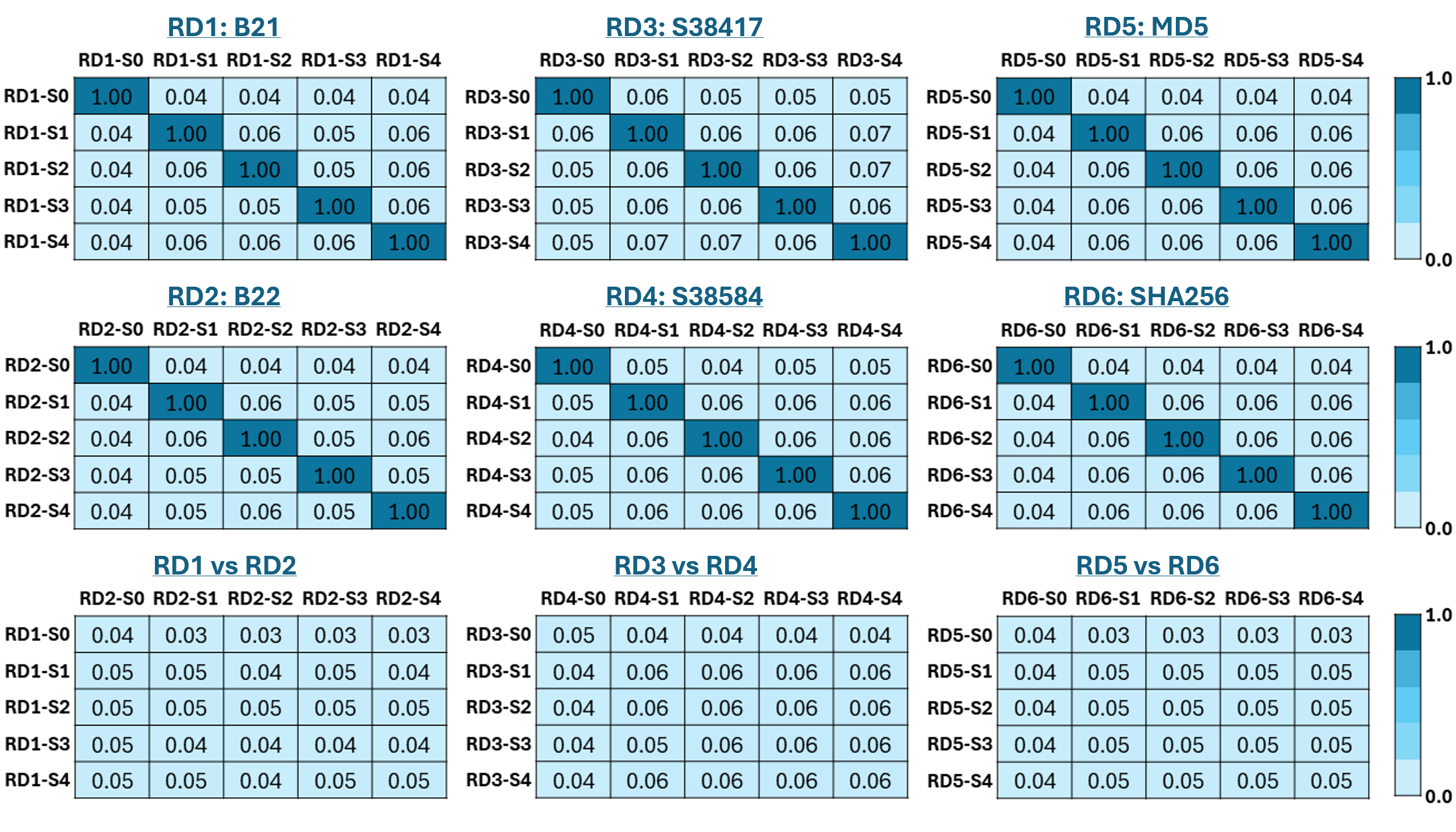}
    \label{fig:heatmap_rd_ciphr}}
    \hfill
    \subfloat[\textit{EvoLUTe} heatmaps.]{\includegraphics[width=0.26\textwidth]{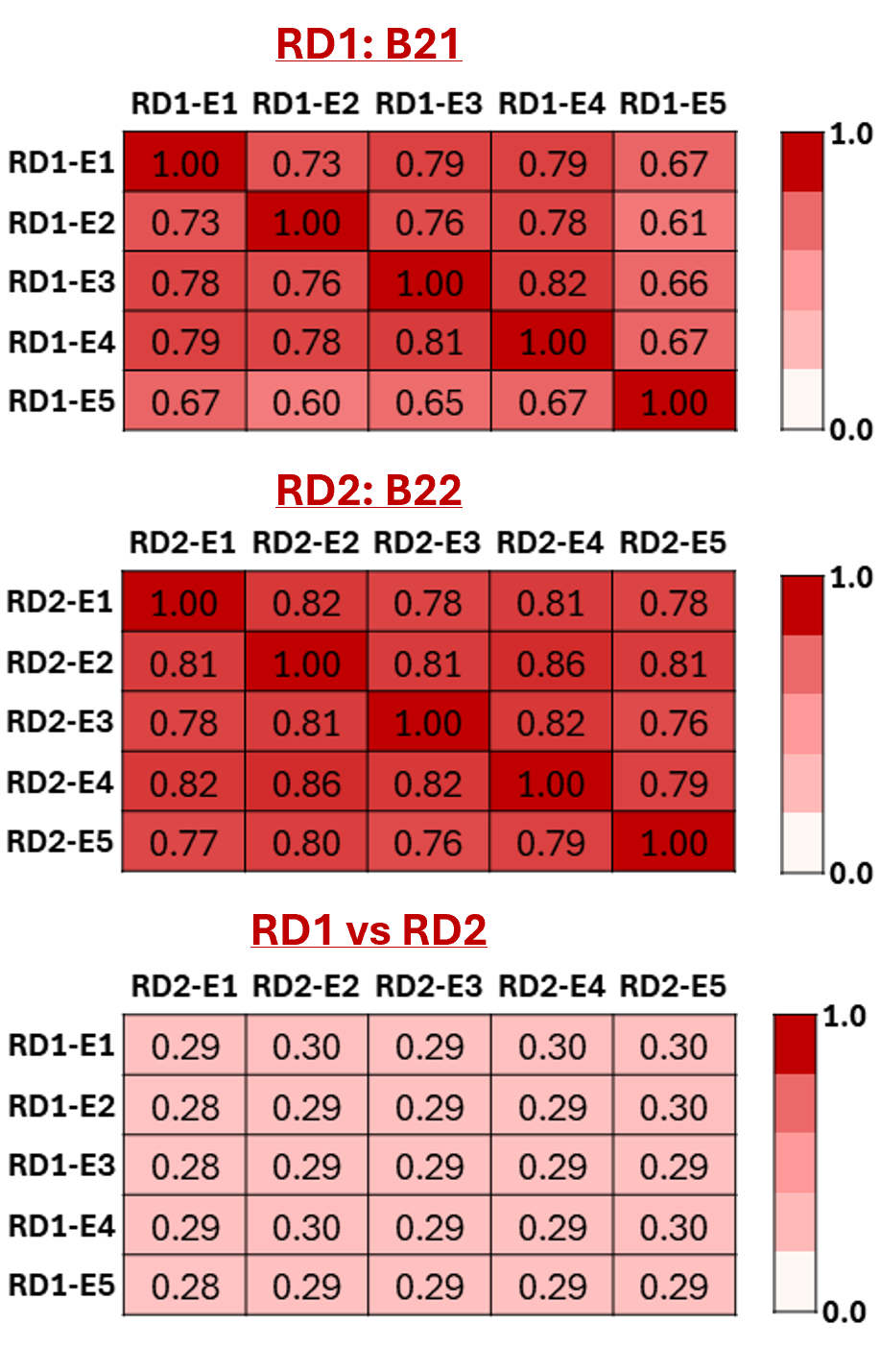}
    \label{fig:heatmap_rd_evolute}}
    \caption{Heatmaps demonstrating \textit{LATTE} results for: (a) \ciphr. (b) \textit{EvoLUTe}.}
    \label{fig:heatmap_rd_tdi_s}
\end{figure*}

\subsubsection{RE using \textit{LATTE}}

For a more extensive analysis of $\mathbf{TDI}$ in \ciphr, we use \textit{LATTE}~\cite{dasgupta2025latte}, a RE attack by a skilled adversary with privileged access to the IC supply chain. Under \textit{LATTE}, the adversary leverages design features and prior knowledge of security countermeasures to generate a library of candidate IPs. The original IP can be recovered from these candidates using the combined functional and structural similarity metrics reported by the LEC tool. Due to the lack of access to the \textit{EvoLUTe} tool and redacted benchmarks, we generated the redacted IPs by emulating the fine-grain redaction technique using an open-source tool \cite{neos}.

Fig. \ref{fig:heatmap_rd_tdi_s} presents the \textit{LATTE} results obtained for \ciphrs and \textit{EvoLUTe}, providing a comprehensive visualization of their performance on various designs. First, the intra-IP variants \textbf{RD$i$-S$k$} generated by \ciphrs are compared to each other, followed by a pairwise inter-IP variant comparison between similar benchmarks. For \textit{EvoLUTe}, we repeat the same intra-IP and inter-IP comparisons for \textbf{RD$1$} (B21) and \textbf{RD$2$} (B22). The heatmaps obtained for \ciphrs (in \textcolor{blue}{\textit{blue}}) are depicted in Fig. \ref{fig:heatmap_rd_ciphr}, showing that the redacted variants have negligible intra-IP and inter-IP similarity scores, implying high degrees of $\mathbf{TDI}_{F}$ and $\mathbf{TDI}_{S}$. 

However, the heatmaps for \textit{EvoLUTe} (in \textcolor{Maroon}{\textit{red}}) redacted variants for the same IPs, depicted in Fig. \ref{fig:heatmap_rd_evolute}, show 10x to 20x higher intra-IP and inter-IP similarity scores compared to \ciphr. The higher values of intra-IP and inter-IP similarity scores indicate significantly more structural and functional similarity, making it more vulnerable to \textit{LATTE}. Hence, we can infer that the redacted IP variants generated by \ciphrs offer enhanced resilience against \textit{LATTE} compared to \textit{EvoLUTe}.

\section{Conclusion and Future Work}
\label{sec:conclude}

In this paper, we have presented an important threat model for hardware IP protection against RE that prevents extraction of design secrets based on a library of known designs. We have demonstrated that the state-of-the-art technologies for IP protection via hardware redaction -- both fine-grain and coarse-grain -- fail to protect against this threat. We have subsequently presented \ciphr, a robust security-aware fine-grain redaction scheme that can effectively protect against this threat by attaining the \textit{indistinguishability} property that many encryption techniques possess. We have provided mathematical analysis, as well as detailed evaluation results based on the software implementation of \ciphrs to prove the robustness and scalability of the proposed IP protection approach. We have shown that \ciphrs approach can seamlessly integrate with commercial ASIC design flow and achieves significantly lower overhead than existing redaction techniques. 

Our future work will include further security evaluation of \ciphrs using the latest \textit{oracle-guided} \cite{Break-and-Unroll_2022, FuncTeller_2023} and \textit{oracle-less} \cite{DANA_2020,gnnunlock,dasgupta2025latte,knowledge_guided_attack} attack vectors under a more comprehensive threat model to demonstrate the true effectiveness of the security-aware randomized transformations proposed in \ciphr. Moreover, we would like to demonstrate the further scalability of the proposed methodology in \ciphrs using large-scale SoC benchmarks and analyze the impact of custom-designed configurable fabrics on the PPA overheads across the various stages of the EDA tool flow. Furthermore, we plan to extend the randomized transformations introduced by \ciphrs to other IP protection techniques and the FPGA design flow.

\bibliographystyle{IEEEtran}
\bibliography{references}

\end{document}